\documentclass[aps,floatfix,twocolumn]{revtex4}

\usepackage{amsmath}
\usepackage{graphicx}
\usepackage{color}
\usepackage{pslatex}
\usepackage{setspace}
\usepackage[normalem]{ulem}

\begin{document}

\title{Time Dependent Quantum Thermodynamics of a Coupled Quantum Oscillator System in a Small Thermal Environment}

\author{George L. Barnes}
\affiliation{Department of Chemistry and Biochemistry,\\Siena College\\Loudonville, NY 12211, USA}
\author{Michael E. Kellman}
\email{kellman@uoregon.edu}
\affiliation{Department of Chemistry and Institute of Theoretical Science, University of Oregon \\ Eugene, OR 97403, USA}
\date{\today}

\begin{abstract}
Simulations are performed of a small quantum system interacting with a quantum environment.  The system consists of various initial states of two harmonic oscillators coupled to give normal modes.  The environment is ``designed" by its level pattern  to have a thermodynamic temperature.  A random coupling causes the system and environment to become entangled in the course of time evolution.  The approach to a Boltzmann distribution is observed, and effective fitted temperatures close to the designed temperature are obtained.  All initial pure states of the system are driven to equilibrium at very similar rates, with quick loss of memory of the initial state.  The time evolution of the von Neumann entropy is calculated as a measure of equilibration and of quantum coherence.  It is argued, contrary to common understanding, that quantum interference and coherence are eliminated only with maximal entropy, which corresponds thermally to infinite temperature.  Implications of our results for the notion of ``classicalizing" behavior in the approach to thermal equilibrium are briefly considered. 
\end{abstract}

\maketitle

\section{Introduction} \label{introduction}

In recent years, the detailed behavior of a very small quantum system interacting with a quantum environment has become a subject of great interest in several fundamental contexts, including the quantum foundations of thermodynamic behavior  \cite{Gemmer:2009,Gemmerarticle}, the nature of the quantum measurement process, and the question of how the classical emerges from the quantum world \cite{Schlosshauer:decoherence}.  In the area of quantum thermodynamics,  Gemmer et al. \cite{Gemmerarticle,Gemmer:2009} have shown in model simulations that thermodynamic behavior emerges in surprisingly small entangled quantum systems.  Microcanonical and thermal behavior have been observed, including the rapid approach to microcanonical equilibrium entropy, and the attainment of a kind of thermal distribution in canonical systems.  In a complementary approach, Popescu et al. \cite{PopescuNature} have shown analytically how thermal behavior arises in ``typical situations."  The systems of Ref. \cite{Gemmerarticle}  are comprised of a few levels with no structure apart from a simple energy level spectrum.  The behavior of systems with real structure in a quasi-thermal environment is a problem of great interest in a number of contexts.  The  time-dependent behavior of select initial states is an issue in fields as seemingly disconnected as small reactive combustion species and large biological light-harvesting machines.  

Our goal in this paper is to perform simulations along the lines of Ref. \cite{Gemmerarticle} for a model system that has sufficient internal structure to have interesting dynamics.   We examine the behavior of a model oscillator system -- a set of normal mode states of two coupled oscillators -- interacting with a finite quantum thermal bath.  We examine the time-dependent quantum behavior of various initially prepared pure states as the oscillator system interacts with the quantum environment.  The total system + environment ``universe'' is in a pure state described by the density matrix $ \rho_{SE}$.  The statistical behavior of the system is described by the reduced density matrix (RDM)  computed as $ \rho_S = Tr_E  \rho_{SE} $.   
In time-dependent simulations of initially pure states of the system interacting with the environmental bath, we find that a Boltzmann distribution is obtained with a meaningful temperature and more or less complete decay of off-diagonal elements, i.e. complete dephasing.  However, there are significant time dependent fluctuations associated with the finite size of the environment.  

Visual insight into the dynamics of the system is obtained by examining spatial density distributions of the system RDM.  A point we emphasize is that, contrary to a commonly held understanding, complete dephasing in the  RDM associated with a thermal distribution does not require the complete absence of quantum interference effects in the system.  This is seen very strikingly in spatial density plots of the system in Boltzmann distributions at various temperatures. 

The outline of the paper is as follows.  In Section \ref{sec:method} we provide an overview of the computational approach for describing a quantum system embedded in a quantum environment.  In Section \ref{sec:results} we describe the results from our simulations while Section \ref{interference} describes the relationship between quantum interference, quantum coherence, and the von Neumann entropy.  In Section \ref{sec:summary} we discuss and summarizes our findings.

\section{Method\label{sec:method}}

In this section we present our model for a system interacting with an environment along with a description of how the dynamical analysis is performed.  We generally follow the approach of Ref. \cite{Gemmerarticle} to simulate a system embedded in a quantum thermal environment, but modify their method as described later in this section for the steps involved in defining the energy levels and the initial state used for time propagation. 

Ref. \cite{Gemmerarticle} devised an environment with a temperature $T$ based on standard statistical mechanical reasoning concerning the degeneracy behavior of the bath energy levels.     The resulting environment yields equilibrated, time-averaged populations obtained from the RDM.  These are found to be in accord with statistical mechanics; i.e. a Boltzmann-like distribution corresponding to $T$ for all initial states of the system in the long time limit.  In this work we will refer to the quantum system as S, the quantum environment as E and the combined system-environment as SE, which taken together constitutes the quantum universe.  

\subsection{System, Environment, Model Hamiltonian, and Temperature \label{sec:method:model}}

We define our total Hamiltonian operator as a sum of three parts

\begin{equation}
\hat{\mathrm{H}} = \hat{\mathrm{H}}_{S} + \hat{\mathrm{H}}_{E} + \hat{\mathrm{H}}_{SE}
\end{equation}

\noindent where $\hat{\mathrm{H}}_{S}$ and $\hat{\mathrm{H}}_{E}$ represent the Hamiltonian of the isolated system and  environment, respectively, and $\hat{\mathrm{H}}_{SE}$ is the system-environment interaction.  We work in the basis of the energy eigenrepresentation of both the system and environment which means that both $\hat{\mathrm{H}}_{S}$ and $\hat{\mathrm{H}}_{E}$  are represented in diagonal form.

\subsubsection{Isolated System Hamiltonian \label{isolatedsystem}}

For the system we take two linearly coupled oscillators, labeled 1 and 2, with Hamiltonian

\begin{equation}   \hat{\mathrm{H}}_{S}^L = (n_1 + n_2) \omega_0 + \kappa (a_1^{\dagger} a_2 + a_1 a_2^{\dagger})  \label{system}  \end{equation}  

\noindent where $\hat{\mathrm{H}}_{S}^L$ is the zero order (non-eigen) S Hamiltonian, $n_1, n_2$ are the numbers of quanta in ``local" modes 1, 2 and $\omega_0, \kappa$ are parameters that we take to be 34.64 and 1.0 in reduced units (the rationale for various parameter choices is detailed below ).  The superscript ``L" indicates the ``local mode" representation.  The coupled system yields normal mode eigenstates that can be labeled by quantum numbers $n_s, n_a$ for the number of quanta in the symmetric and antisymmetric  modes.  The Hamiltonian in this normal mode representation is

\begin{equation}  \hat {\mathrm{H}}_S = n_s \omega_s + n_a \omega_a = n_s  (\omega_0 - \frac{1}{2} \kappa) + (\omega_0 + \frac{1}{2} \kappa)     \end{equation}  

\begin{figure}[htb]
\rotatebox{270}{\scalebox{.3}{\includegraphics{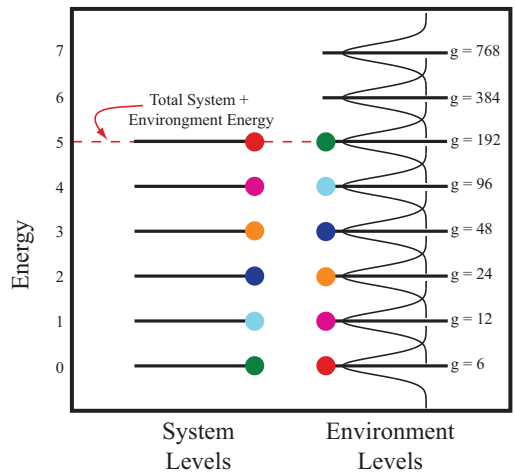}}}
\caption{The energy level diagram for both the system and environment.  There are 6 system eigenlevels and 8 environment energy levels.  The degeneracy scheme for the environment is shown along with the gaussian spread.   We select initial states of the system and environment making use of the notion of a constant total universe energy (dotted red line) with the initial system state thought of as a fluctuation within the SE universe, enabled by the environment bath.  For a given fluctuated system state (colored dot on system level) the corresponding environment state (matching colored dot on environment level) is defined in order to conserve the total universe energy.\label{fig:ELD}}
\end{figure}

\noindent The coupling in Eq. \ref{system} preserves the total quantum number $N = n_1 + n_2 = n_s + n_a$, often referred to as the polyad number.  Associated with a given $N$ are a set of $N+1$ normal mode states, referred to as a polyad of states.  Each distinct polyad constitutes an isolated system since the Hamiltonian preserves the polyad number (i.e. distinct polyads are not coupled).  The normal mode energies are equally spaced within a polyad.  In contrast to normal mode states $| n_s, n_a \rangle$, which have good quantum numbers $n_s, n_a$ and a trivial time dependence, local mode states that initially have quantum numbers $n_1, n_2$ evolve in time in an oscillatory manner.  For example, an initial local mode overtone state $ |n_1, n_2 \rangle = | N, 0 \rangle$ oscillates back and forth with the partner overtone state $|0, N \rangle $ because of the coupling in Eq. \ref{system}.  More details about such systems are available in the molecular literature \cite{KellmanAnnRev,Xiao89sphere,Xiao90cat,Svitak95}.  We will examine the time dependent behavior of both normal mode and local mode overtone states interacting with E.  For S we take the polyad of levels with  $N = 5$, a set of six equally spaced levels  as shown in Figure \ref{fig:ELD}.  We work in reduced units in this paper and hence  we can define the value of the spacing between states within the $N = 5$ polyad as 1.  Although the calculations are performed in reduced units,  we present final results in wavenumbers and picoseconds based on the absolute spacing of 111.77 cm$^{-1}$ between these polyad states.  This value  corresponds to the parameter $\kappa$ in Eq. \ref{system} in absolute units  and was adapted from a fit to the water stretching mode spectrum in Ref. \cite{Xiao89sphere,Xiao90cat}.    For convenience we will label each energy eigenvalue of the normal mode system using the quantum number $n = 0 \cdots 5$.

\subsubsection{Isolated Environment Hamiltonian}     \label{isolatedenvironment}

In treating the quantum environment E we make as straightforward an approximation as possible.  Namely a set of evenly spaced harmonic levels that can be completely defined using an energy quantum number $m$ and a degeneracy quantum number $l$.  These quantum numbers together specify each zero order E state  as $|m,l\rangle$.  (The zero-order harmonic degeneracy of E will be broken later in Eq. \ref{H_E}.)  The zero order E Hamiltonian is defined as  

\begin{equation}
\hat{\mathrm{H}}_{E}^0 |m,l\rangle = m \omega_{E} |m,l\rangle
\end{equation}

\noindent where $\omega_{E}$ is a parameter of the model.  In this work we choose to make the zero order harmonic spacing of E equal to that of the system.  This is based on the idea that a real E will have a continuum of levels and the system levels will interact most strongly with those levels closest in energy.  The degeneracy of each level is given by $g(m)=Ab^{m\omega_E}$ where $A$ and $b$ are  parameters of the model; the rationale for this degeneracy behavior is explained shortly.  The zero order E energy levels are  shown in Figure \ref{fig:ELD}.   

Ref. \cite{Gemmerarticle}  made use of this type of degeneracy scheme as a way to approximate the true degeneracy scheme that would be found for a large E, e.g. a large collection of oscillators.   At this point it is instructive to examine what temperature such an E energy level pattern should represent.  The thermodynamic definition of temperature is given by

\begin{equation}
\frac{1}{T} = \frac{\partial S}{\partial U}.
\end{equation}

\noindent The connection with statistical mechanics is made through the relation $S = k_b \ln W$, and then related to our construction by using the degeneracy formula  $S = k_b \ln\left ( Ab^U\right)$ where $k_b$ is the Boltzmann factor.  This yields the expression 

\begin{equation}
T = \frac{1}{k_b\ln\left( b \right )}\label{eq:Temp}
\end{equation}

\noindent which shows that the base of the exponential scaling of the degeneracy in fact defines the temperature.  In this work we take $b = 2$ which leads to a temperature of 230.41 degrees Kelvin.  We assume that we work in a sufficiently narrow energy range of the bath that $T$ is energy independent.

The SE interaction will be taken to be a random coupling, as we now describe.    Following Ref. \cite{Gemmerarticle}, the SE interaction will disrupt the perfect degeneracy pattern defined above.  We chose to include the diagonal portion of the SE interaction directly in the E Hamiltonian and represent it as

\begin{eqnarray}
\hat{\mathrm{H}}_{E}^{shift}|m,l\rangle &=& X(m,l) |m,l\rangle\\
\hat{\mathrm{H}}_{E} &=& \hat{\mathrm{H}}_{E}^0 + \hat{\mathrm{H}}_{E}^{shift}  \label{H_E}
\end{eqnarray}

\noindent where $X(m,l)$ is a random variate selected from a gaussian distribution with zero mean and standard deviation, $\sigma=\alpha\omega_{bath}\sqrt{2}$.  The eigenvalue equation for the final E Hamiltonian is given by

\begin{equation}
\hat{\mathrm{H}}_{E} |m,l\rangle = \left [n\omega_{E} + X(m,l) \right ] |m,l\rangle
\end{equation}

\noindent This spreading of the degenerate environment eigenvalues is shown in Figure \ref{fig:ELD} by the gaussians centered on each level.

\subsubsection{System-Environment Interaction}

 We now define our composite system states as $|n\rangle \otimes |m,l\rangle \equiv |n,m,l\rangle$ and turn our attention to the off-diagonal SE interaction terms.  Since we cannot presume any particular form of the interaction, all off-diagonal elements are in general non-zero and given by

\begin{equation}
\langle n,m,l | \hat{\mathrm{H}}_{SE} | n^\prime, m^\prime, l^\prime \rangle = Y(n,m,l)
\end{equation}

\noindent where $Y(n,m,l)$ is a random variate selected from a gaussian distribution with zero mean and standard deviation $\sigma=\alpha\omega_{bath}$.   (Note that although $\alpha$ is the same, this is a different standard deviation than above.)  Due to the random nature of the $Y$'s, in general this would produce a non-Hermitian representation.  To avoid this problem, we only generate random variates for the upper triangle and map these to the lower triangle.  The resulting Hamiltonian is diagonalized to yield the energy eigenvalues and vectors of SE, which allows for analytic time propagation of our initially selected states.  

\subsubsection{Initial State Selection}

Up to this point we have followed Ref. \cite{Gemmerarticle} closely, with only slight deviations from their approach.  They showed that for one specific initial S state  $| \psi \rangle$ combined with one initial E state $| \epsilon \rangle$  (taken together as $| \psi \rangle | \epsilon \rangle$), the equilibrated populations of S energy levels obtained from the RDM were well described by the Boltzmann distribution, i.e. an exponential distribution with a fitted temperature $T_{fit}$ close to a suitably defined analytic $T$.    We tried a straightforward adaptation of this method for different initial S states by keeping the same initial E or ``bath" state  and found greatly varying $T_{fit}$ values depending on that initial choice.  This is not the way a bath should behave.  

To avoid this unfavorable result, we adopt a point of view vividly expressed in a 1910 lecture of Einstein \cite{Damour:2006}  on connections between the classical microcanonical and canonical ensembles, the Boltzmann factor $e^{-\beta E}$, and fluctuations of a Brownian particle in a gravitational field.  The Brownian particle ``system" moves from one fluctuated state (i.e. vertical displacement) to another, with the total SE state conserving energy, leading to the Boltzmann probability factor $e^{-\beta E}$ when the accompanying entropy fluctuations are taken into account.  Averaged over time, the Brownian particle is characterized by the canonical distribution.  Along analogous lines, we want to think of the system S embedded in the environment E as undergoing canonical (i.e. thermal) fluctuations in a universe SE of fixed total energy $E = 5$, which in classical statistical mechanics would therefore be described by the microcanonical ensemble.  In terms of the notation introduced above, we use various initial ``fluctuated" states $|n\rangle$ of the system with   $n = 0 \cdots 5$, with corresponding SE basis states $|n\rangle \otimes |m,l\rangle \equiv |n,m,l\rangle$ defined earlier that have the zero order energy $E$.  We choose the initial state of SE to be a pure superposition state of the  $|n,m,l\rangle$ basis sates, with equal coefficients for each E basis state.

We do this not because this particular initial state is necessarily more likely than others, but simply as a reflection of our ignorance about the condition of the environment in the initial state.  The equal coefficient assumption may be said to correspond to ``microcanonical" conditions of the environment in the initial fluctuated system state.  

We illustrate the balance between the total universe energy with the portion placed in S and E in Figure \ref{fig:ELD}.  For example, if we have a total universe energy of 5 and we wish to place the system initially in the third S energy level ($n=2$, dark blue dot), then in order to maintain the total energy of the universe E must occupy the fourth environment energy level ($m=3$, dark blue dot). All of the quasi-degenerate levels for the fourth E have equal initial probability, which is equivalent to supposing that we are starting our state on a particular energy ``micro-shell'' given our fluctuated S state.   The choice of the total energy in the universe is arbitrary; by picking 5 we are able to excite all levels within S as illustarted by the series of colored circles in Figure \ref{fig:ELD}.  Such an adaptation of the procedure of Ref. \cite{Gemmerarticle}, along the lines of Einstein's fluctuating system, will be shown in to yield good results for $T_{fit}$ for all initial states considered.  In this work we made use of 6 S eigenlevels and 8 E eigenlevels with the degeneracy scheme noted in Figure \ref{fig:ELD}.  This yields 1530 E levels and a total of 9180 SE levels.  We note that more E energy levels  were included than system levels (8 vs. 6, neglecting degeneracy).  This choice yielded more consistent results for the fitted temperature obtained for each initial state.  One way of thinking about this result is that the ``extra'' E energy levels are needed to converge the calculation.

\subsection{Dynamical Analysis}

\subsubsection{Reduced Density Matrix and Entropy}

In our analysis we perform analytic time propagation of the initial states described above to obtain the time dependent reduced density matrix.  We use the notation $|\Psi(t)\rangle$ to specify our time propagated SE composite.  We begin with the universe density operator $\rho_{SE}$ and define \cite{nielsenchuang} the RDM as

\begin{equation}
\rho_S=Tr_E \rho_{SE}.
\end{equation}

\noindent The elements of the RDM are then given by

\begin{equation}
\rho^{n,n^\prime}_{S} = \sum_{m,l} \langle n,m,l|\Psi(t)\rangle\langle\Psi(t)|n^\prime,m,l\rangle
\end{equation}

\noindent  where the summation serves to give the partial trace over E.  In general the RDM is a complex, non-diagonal Hermitian matrix.  The diagonal elements represent ``populations'' of the eigenstates of S, while the off-diagonal elements represent ``coherences'' between the respective S eigenstates.  Our main focus in this work is not on the decoherence process, though our simulations do include this effect.  See Section \ref{interference} for a discussion of coherence and quantum interference in systems with a finite temperature.

We will use the RDM to obtain both the von Neumann entropy \cite{nielsenchuang} and the time dependent spatial densities.  We first diagonalize the RDM to find its eigenstate basis $\{ |\phi_i\rangle \}$.  By definition the RDM in this basis is given by

\begin{equation}
\rho^{i,j}_S = \mathrm{P}_i \delta_{i,j}
\end{equation}

\noindent where $\mathrm{P}_i$ are the eigenvalues of $\rho^{n,n^\prime}_S$ as well as the populations associated with the basis  $ \{|\phi_i \rangle \}$.  The von Neumann entropy is defined using these populations as

\begin{equation}
S_{vN}=\sum_iP_i{\ln P_i}     \label{vNentropy}
\end{equation}

\noindent The von Neumann entropy of a pure state is zero while the maximal entropy is obtained when all states are equally likely.  

\noindent The total spatial density at any instant in time is given by 

\begin{equation}
\left | \langle x|\Psi\rangle \right |^2 = \sum_i \mathrm{P}_i \left | \langle x|\phi_i\rangle \right |^2
\end{equation}

\noindent We emphasize that the $|\phi_i\rangle$ are in general time-dependent linear combinations of the system eigenstates.  By calculating the spatial density as a function of time, insight into the dynamics is obtained.

\subsubsection{Simulation Temperature} 

In Section \ref{sec:method:model} we presented an analytic expression for the temperature based on an E degeneracy formula.  However, the true levels do not precisely follow this degeneracy formula due to the SE interaction.  We therefore define our temperature for each initial state through a fitting procedure.  For a given temperature the Boltzmann distribution defines the expected populations for each S eigenlevel.  After our initially prepared state reaches equilibrium, the RDM has dephased and hence approximately yields these populations.  Fitting the populations obtained from the simulation to the Boltzmann distribution yields the temperature.  However, the populations from the RDM at any instant in time would yield poor results since there are substantial fluctuations in our finite universe.  We therefore calculate and fit time averaged populations. Specifically we take one hundred samples in a 2 picosecond window  to form our time averages.  We note that any fluctuations that take place on a longer time scale will be missed by this approach and hence difference choices of the time window will change the fitted temperature. 

\section{Results\label{sec:results}}

We want to compare the dynamics and final equilibrium distributions obtained from simulations of several different initial states.  Each initial state is a product of  pure S state $| \psi \rangle$ and E state $| \epsilon \rangle$; each S (E) pure state is in general a superposition of S (E) eigenstates.  The initial S state is chosen to be either an energy eigenstate of the isolated normal mode system or the time-dependent local mode state described above in Section \ref{isolatedsystem};  the corresponding E state is selected as  described in Section \ref{isolatedenvironment}.  The initial product SE state $| \psi \rangle | \epsilon \rangle$  becomes entangled through interaction between S and E.    

\subsection{Time dependent approach to Boltzmann thermal equilibrium}  

Our first inquiry is to determine to what extent each of the initial states arrives at the same set of S populations at equilibrium.  Visual inspection of the final equilibrated S populations in the RDM  for each initial state shows that they all behave qualitatively  in a Boltzmann-like manner with $P_0>P_1>\cdots>P_5$.  

\subsubsection{Time-dependent von Neumann entropy}

\begin{figure}[htb]
\rotatebox{270}{\scalebox{.3}{\includegraphics{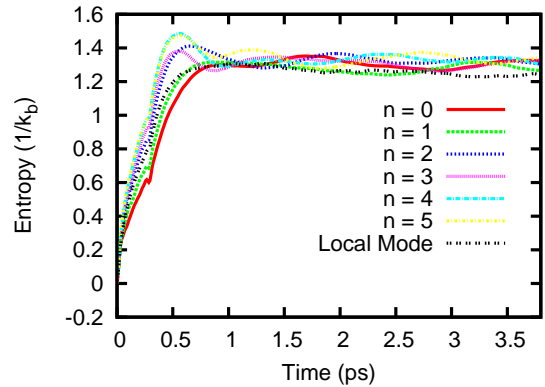}}}
\caption{Time dependent entropy for various initial states.\label{fig:entropy}}
\end{figure}

Below we show that in fact these populations are well fit to a Boltzmann distribution.  Here we first make use of the time dependent dynamics of the von Neumann entropy as a convenient way to compare the final population sets for each of these initial states using a single curve for each initial state rather than six curves for each initial state.  The entropy is calculated by diagonalizing the RDM to obtain a set of populations and computing $S_{vN} = \sum_iP_i\ln\left(P_i\right)$.  If S has undergone complete dephasing then the off-diagonal elements of the RDM are zero and hence the RDM is already diagonal so that the populations used to calculate the entropy are the populations of  S eigenlevels.  Hence if two states reach approximately the same final entropy it can safely be assumed that they have similar S populations if these states act in a Boltzmann manner.  Of course we do not have perfect dephasing in our finite universe, so the time-varying off-diagonal elements will cause fluctuations of the entropy.  Figure \ref{fig:entropy} displays the time evolution of the entropy for our initial states.  Since each state is initially a pure state, it has zero von Neumann entropy by definition.  All of these states reach the same asymptotic entropy of approximately 1.3 with much of the increase happening within the first picosecond of the simulation.  After that time, we observe the expected fluctuations in the entropy due to the finite size of both S and E.  The maximal entropy for this system of 6 levels in the $N = 5$ polyad corresponds to $ \ln\left ( 6 \right ) \approx 1.79$, meaning that this particular E yields an equilibrium S state that corresponds to about 73\% of the maximal entropy.

\subsubsection{Boltzmann thermal distribution and temperature}

\begin{figure}[htb]
\rotatebox{270}{\scalebox{.3}{\includegraphics{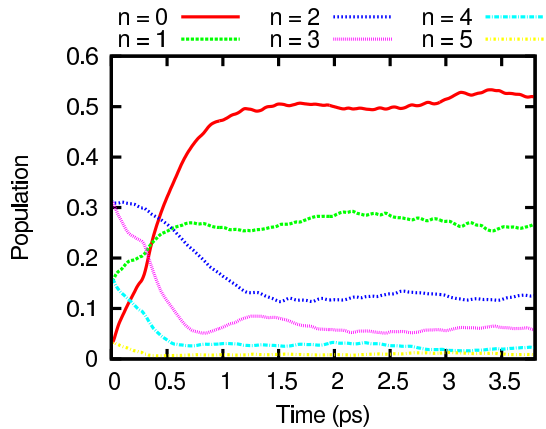}}}
\caption{Time dependent populations for the system eigenstates.  The initial system state is the local mode.\label{fig:pop}}
\end{figure}

\begin{figure}[htb]
\rotatebox{270}{\scalebox{.33}{\includegraphics{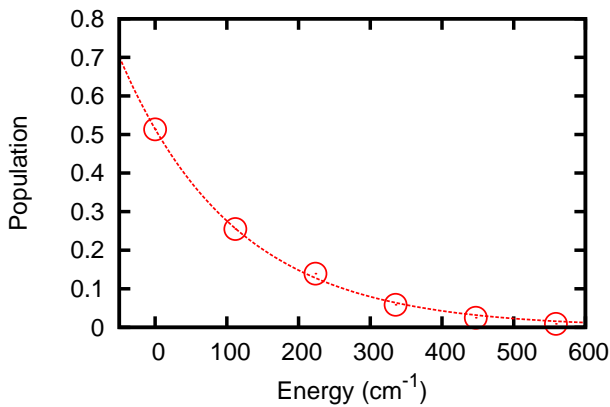}}}
\caption{The above plot displays an example of the time averaged populations obtained from the simulations for a time window after equilibration.  The dashed line is the best fit to the Boltzmann distribution.  The population data is well represented by a Boltzmann distribution.\label{fig:Bfit}}    
\end{figure}

The equilibrium value for the percentage of maximal entropy obtained is one measure of temperature in a Boltzmann-like system, since maximal entropy can only be obtained if all eigenlevels within S have equal probability, i.e. only at infinite temperature.  Hence Figure \ref{fig:entropy} illustrates that each state considered here may be reaching approximately the same temperature, as hoped.  A  direct measure of the temperature is obtainable by looking explicitly at the populations of S eigenlevels.  For example in Figure \ref{fig:pop} we show the time evolution of the population for the local mode initial state.    By taking a time window    from such a set of curves an average population is obtained and subsequently fit  to an exponential (Boltzmann) distribution, as shown in Figure \ref{fig:Bfit}.  This particular curve corresponds to a temperature of 231.3 K with a standard error obtained from the non-linear fit of 5.7 K.    Note that this uncertainty does not correspond to the magnitude of the fluctuations, only the uncertainty in the fit parameters for a given set of averaged populations.  The temperature obtained from the fit will depend on those average populations and hence on the time window selected as well as its size.  This particular time window happens to yield a temperature very close to that obtained from Eq. \ref{eq:Temp}.  We present, in Table \ref{tab:temp}, the fitted $T_{fit}$ for all states during two different time windows.  The average temperature from all states and time windows is 239 K with a standard deviation of 13 K.  The largest fitted temperature difference between the two time windows for a given S state is 44 K.  The expected temperature of 230.41 K, obtained from Eq. \ref{eq:Temp}, is well within one standard deviation of our average value 239 K.  The fitted values are also consistent among themselves, with ten of the fourteen temperatures falling within one standard deviation of the average value, and all but one falling within two standard deviations.  Overall these results are in good agreement with the expected temperature from Equation \ref{eq:Temp}.  Better results could be obtained by lengthening the time window, thereby averaging out fluctuations, as is evident even from averaging the temperatures from the two windows in Table \ref{tab:temp}.  

\begin{table*}[htb]
\caption{A comparison of the fitted $T_{fit}$ in degrees Kelvin for all initial states for two different time windows.  The average and standard deviation of all $T_{fit}$ are 239 and 13, respectively.\label{tab:temp}}
\begin{tabular}{cccccccc}
\hline
 & $n=0$ & $n=1$&$n=2$& $n=3$& $n=4$& $n=5$& LM$^a$ \\
\hline\hline
t=3.8-5.8 ps$^b$ & ~236 $\pm$ 10.$^c$~ & ~234 $\pm$ 6.5~ & ~225 $\pm$ 14.~ & ~259 $\pm$ 7.0~ & ~225 $\pm$ 7.0~ & ~235 $\pm$ 8.0~ & ~246 $\pm$ 7.0~ \\
t=7.6-9.6 ps & 229 $\pm$ 3.5 & 250 $\pm$ 2.8 & 238 $\pm$ 5.3 & 231 $\pm$ 6.1 & 269 $\pm$ 7.4 & 237 $\pm$ 2.6 & 231 $\pm$ 5.7 \\
\hline
\multicolumn{8}{l}{$^a$ Local Mode State} \\
\multicolumn{8}{l}{$^b$ Each time window contains 100 samples} \\
\multicolumn{8}{l}{$^c$ Uncertainty given is the asymptotic standard error of the fitted $T_{fit}$.}
\end{tabular}
\end{table*}

\subsection{Reduced density matrix and time dependent spatial dynamics}  

\begin{figure*}[htb]
\begin{tabular}{ccc}
\rotatebox{270}{\scalebox{.28}{\includegraphics{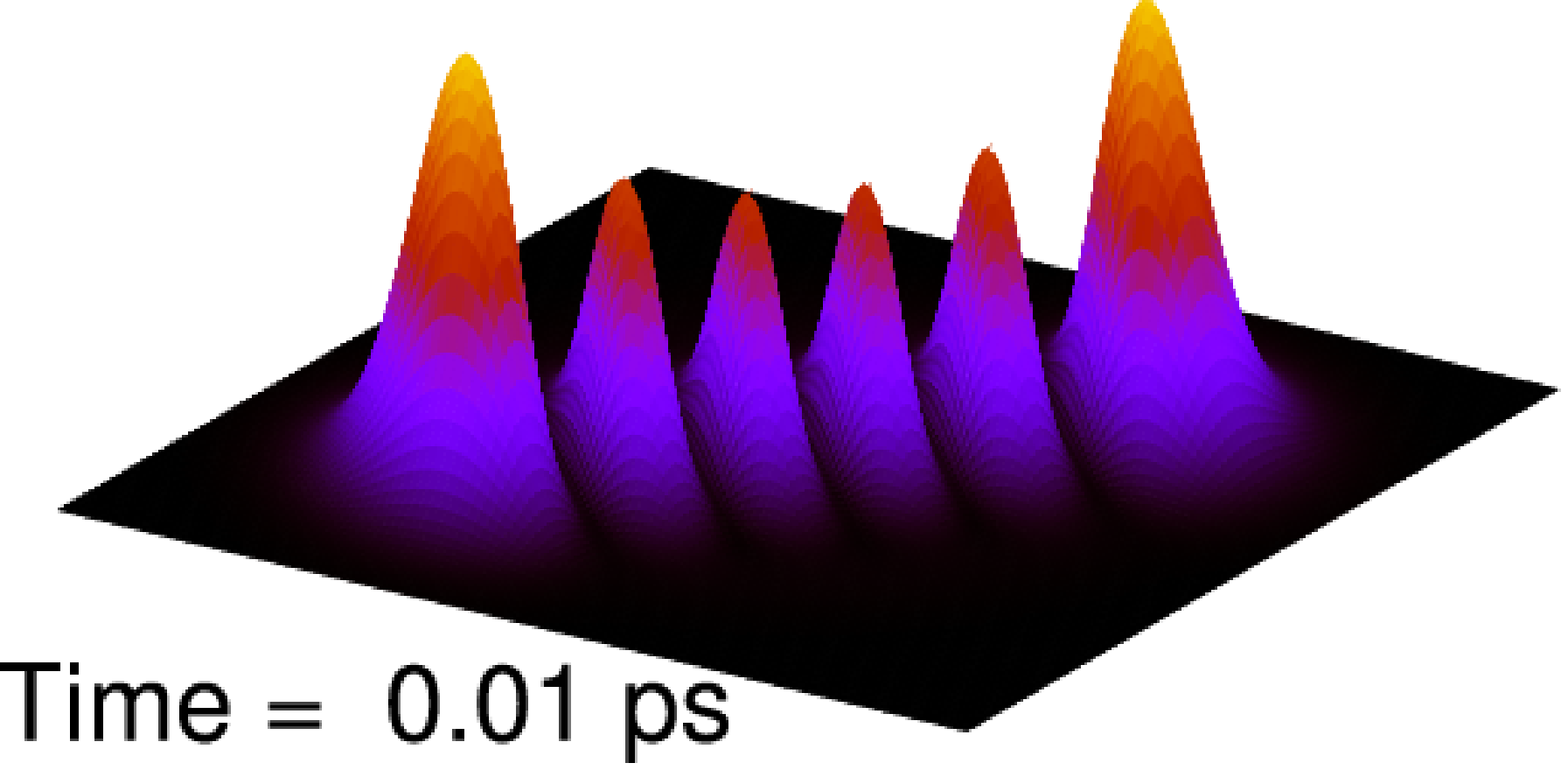}}} &
\rotatebox{270}{\scalebox{.28}{\includegraphics{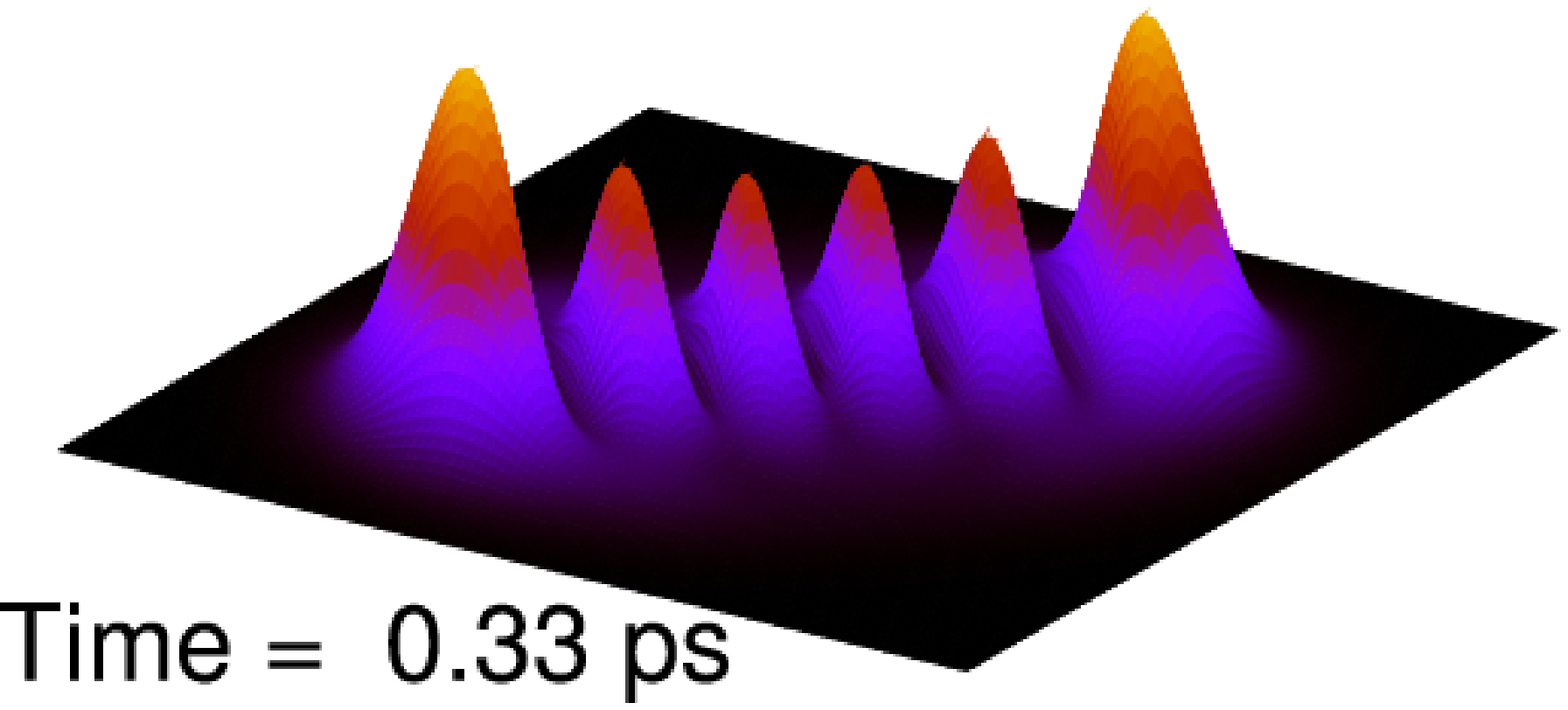}}} &
\rotatebox{270}{\scalebox{.28}{\includegraphics{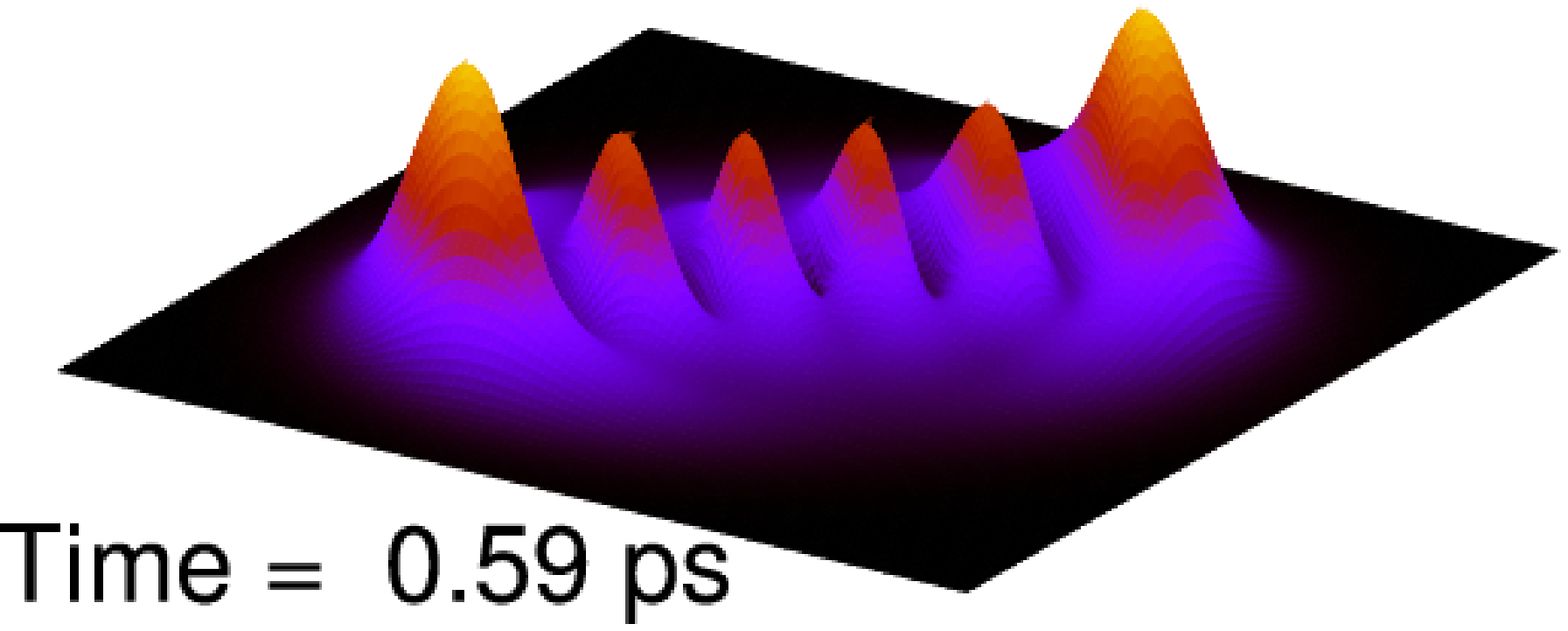}}} \\

\rotatebox{270}{\scalebox{.28}{\includegraphics{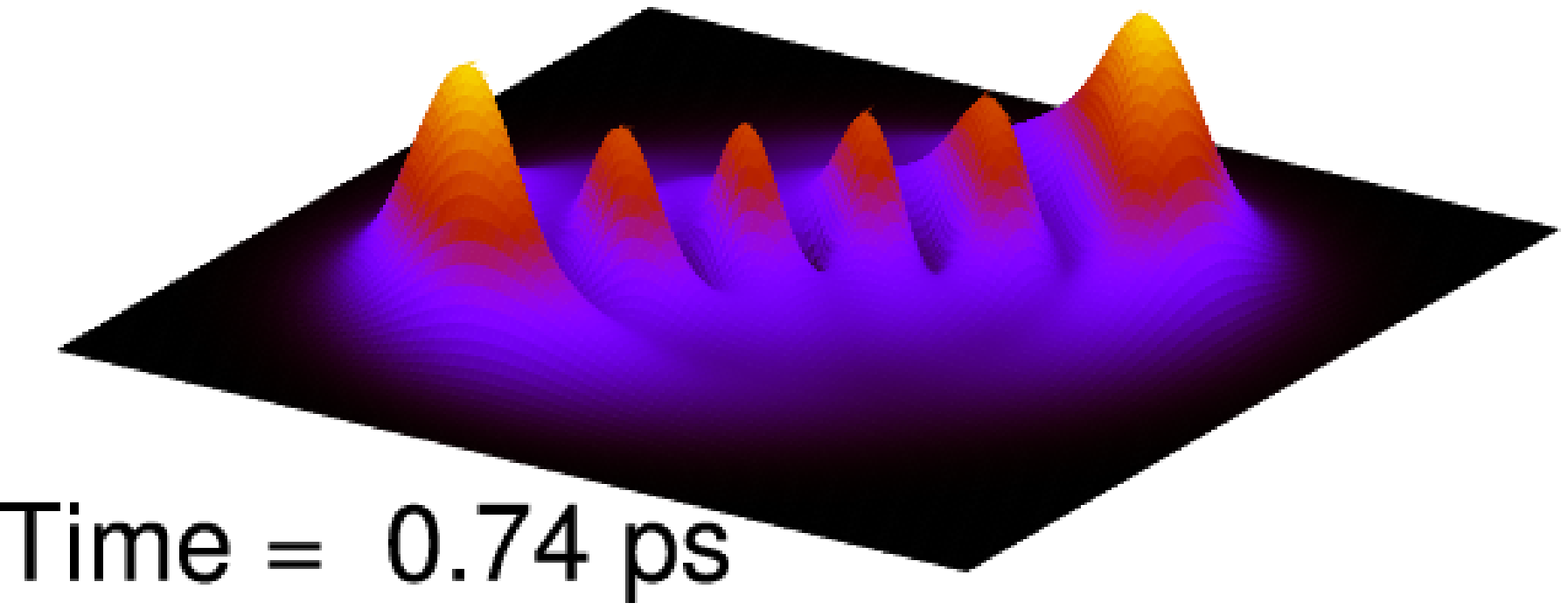}}} &
\rotatebox{270}{\scalebox{.28}{\includegraphics{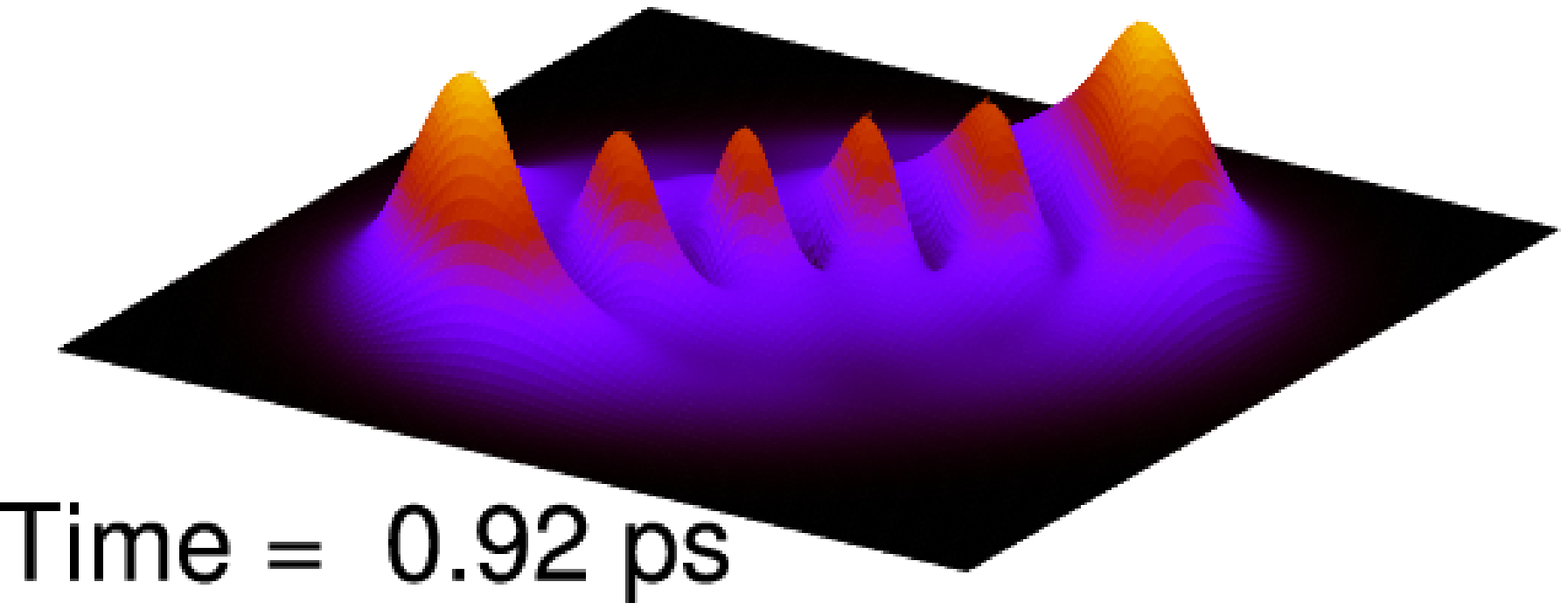}}} &
\rotatebox{270}{\scalebox{.28}{\includegraphics{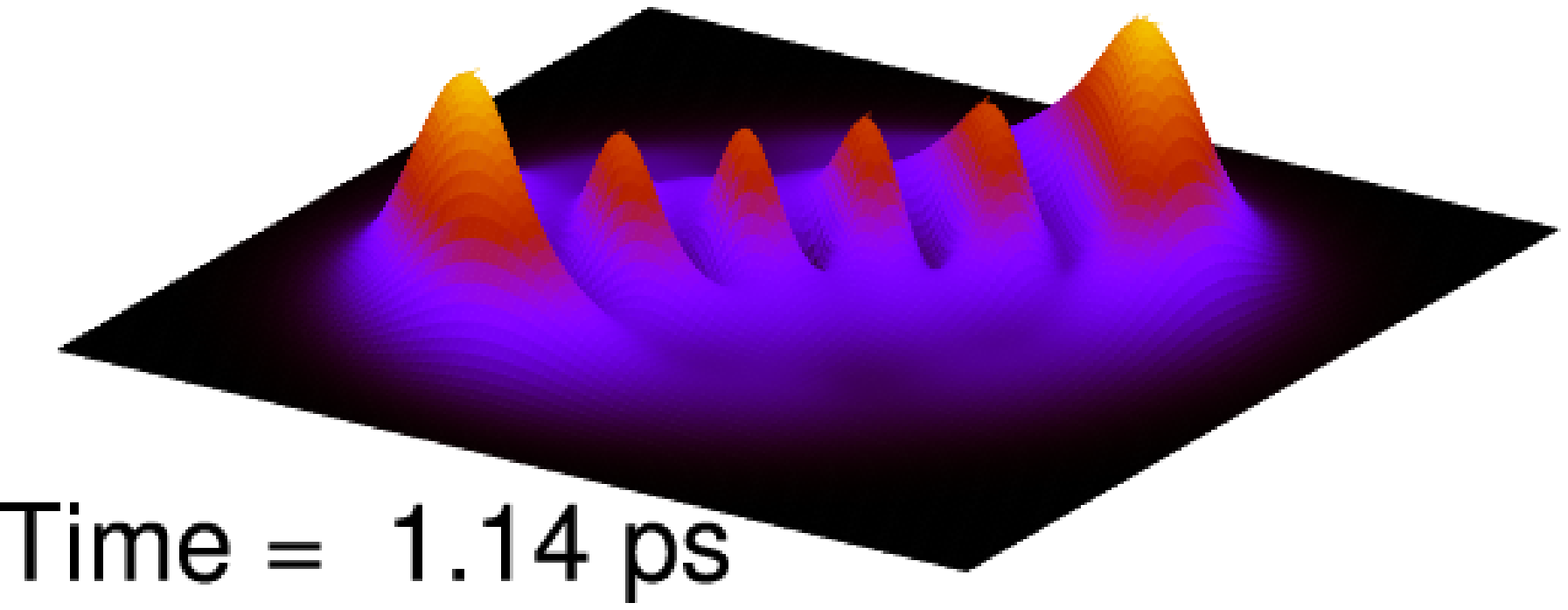}}} \\
\end{tabular}
\caption{Snap-shots of the time evolution of the total state density for the first level of the $N=5$ polyad.\label{fig:NSL1}}
\end{figure*}

\begin{figure*}[htb]
\begin{tabular}{ccc}
\rotatebox{270}{\scalebox{.28}{\includegraphics{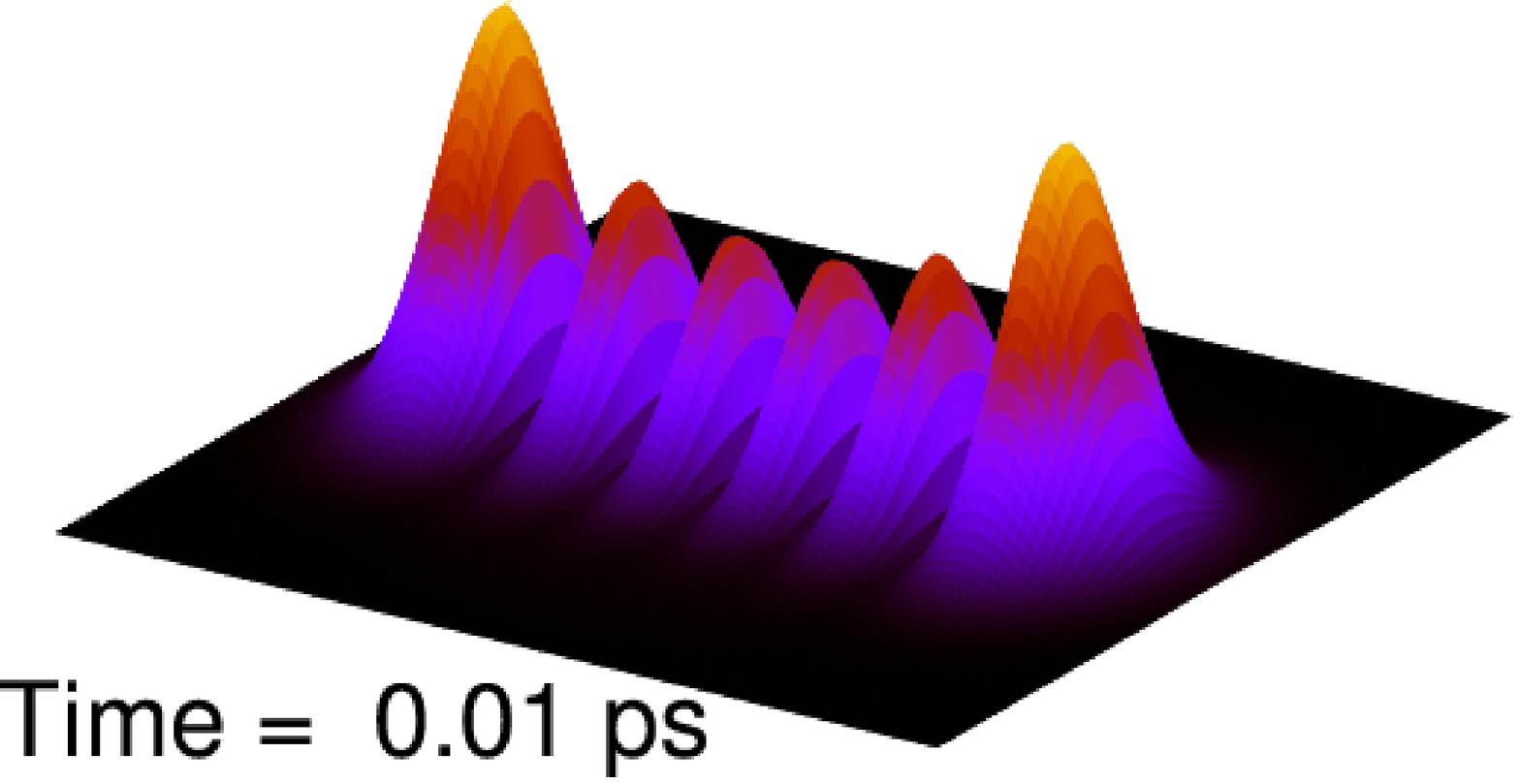}}} &
\rotatebox{270}{\scalebox{.28}{\includegraphics{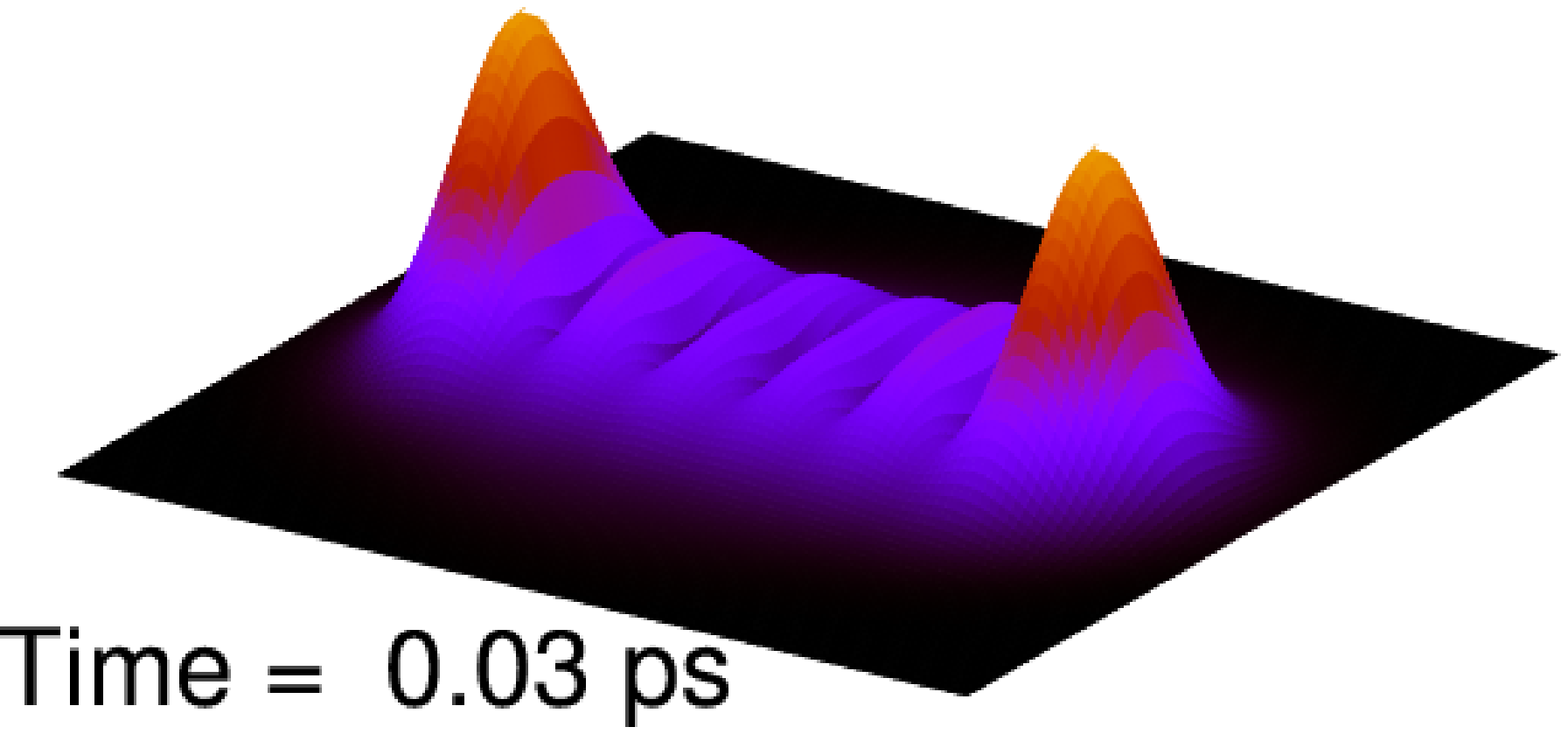}}} &
\rotatebox{270}{\scalebox{.28}{\includegraphics{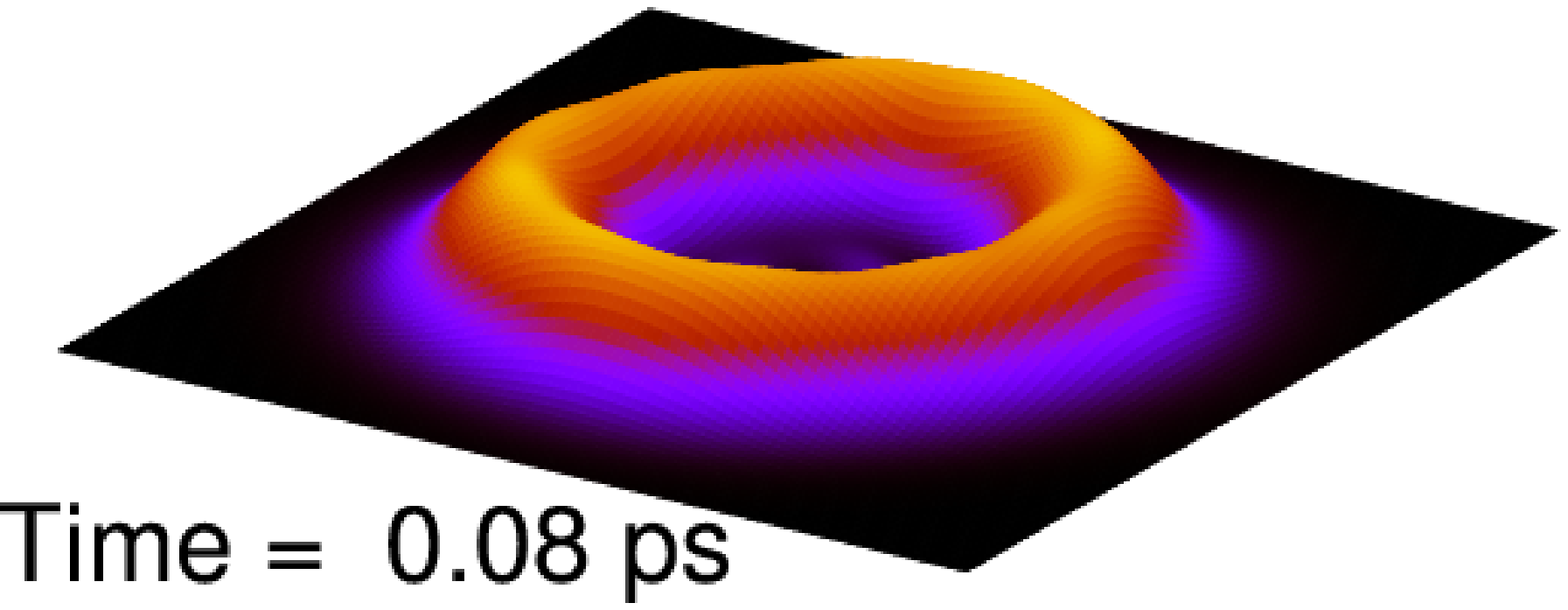}}} \\

\rotatebox{270}{\scalebox{.28}{\includegraphics{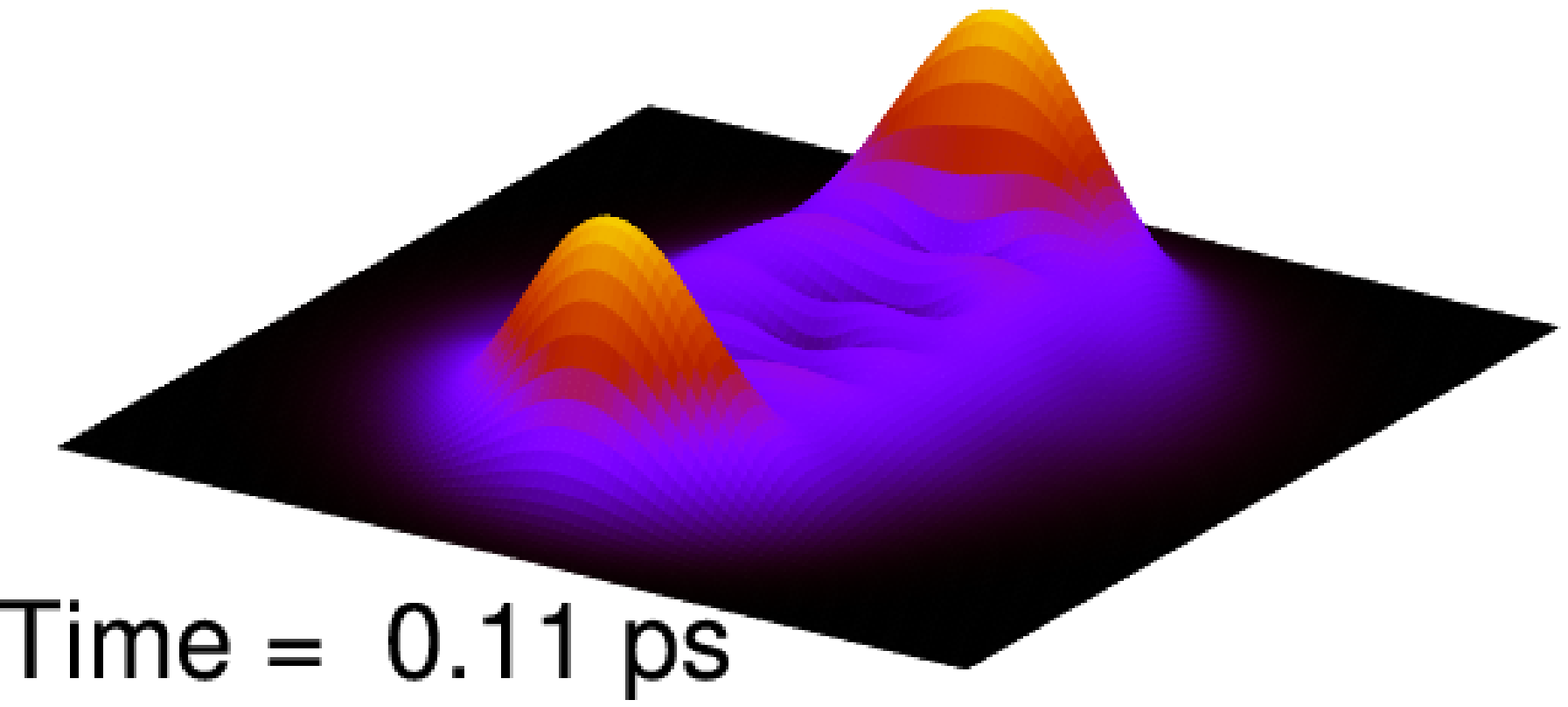}}} &
\rotatebox{270}{\scalebox{.28}{\includegraphics{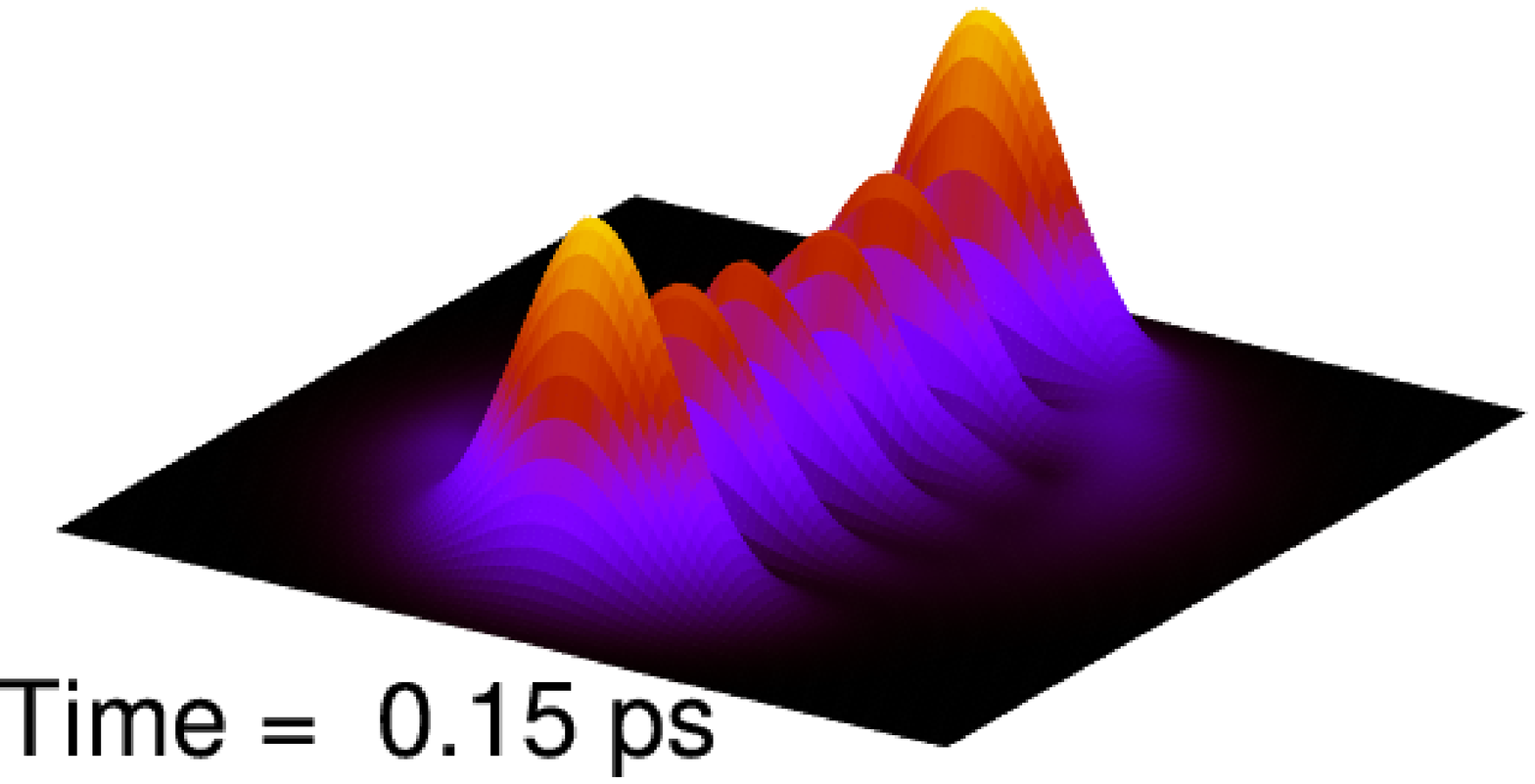}}} &
\rotatebox{270}{\scalebox{.28}{\includegraphics{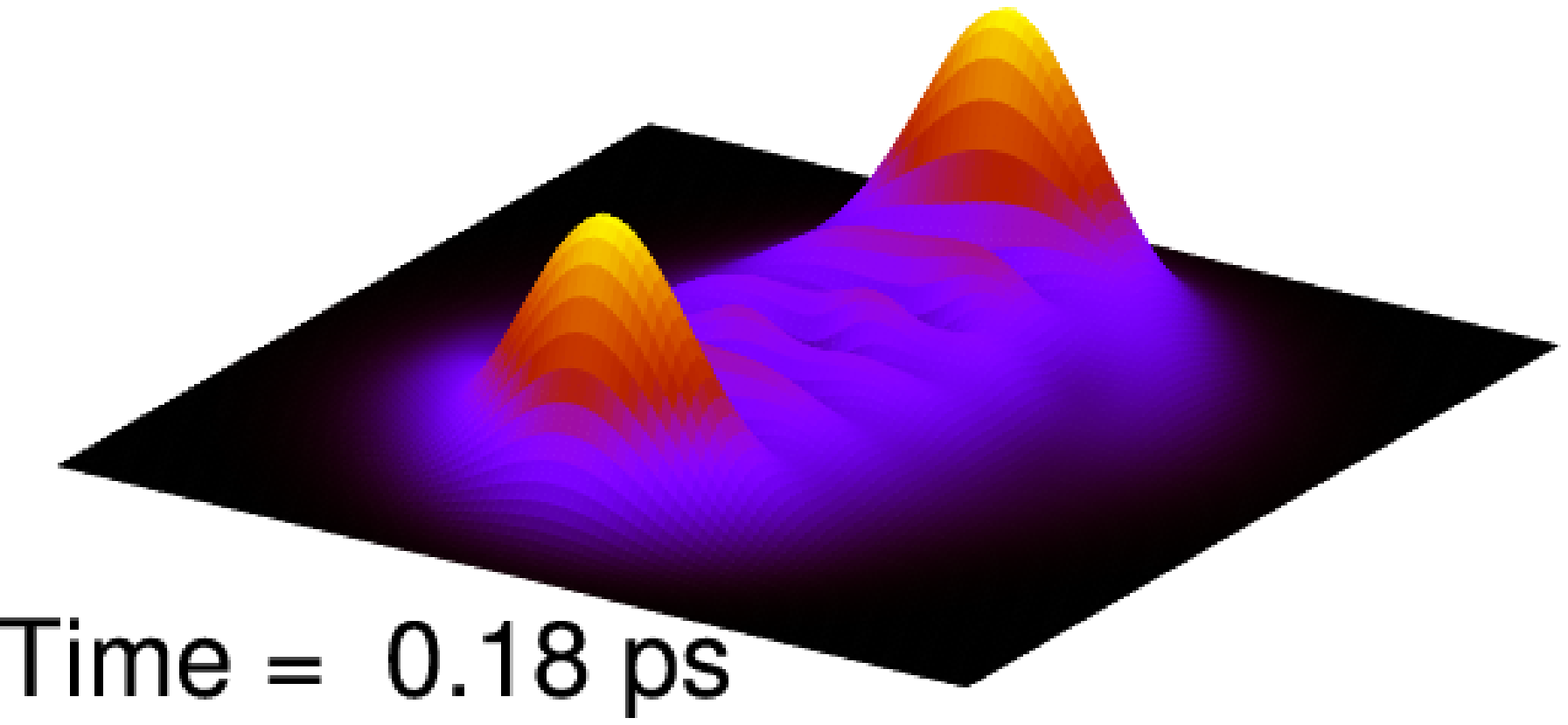}}} \\

\rotatebox{270}{\scalebox{.28}{\includegraphics{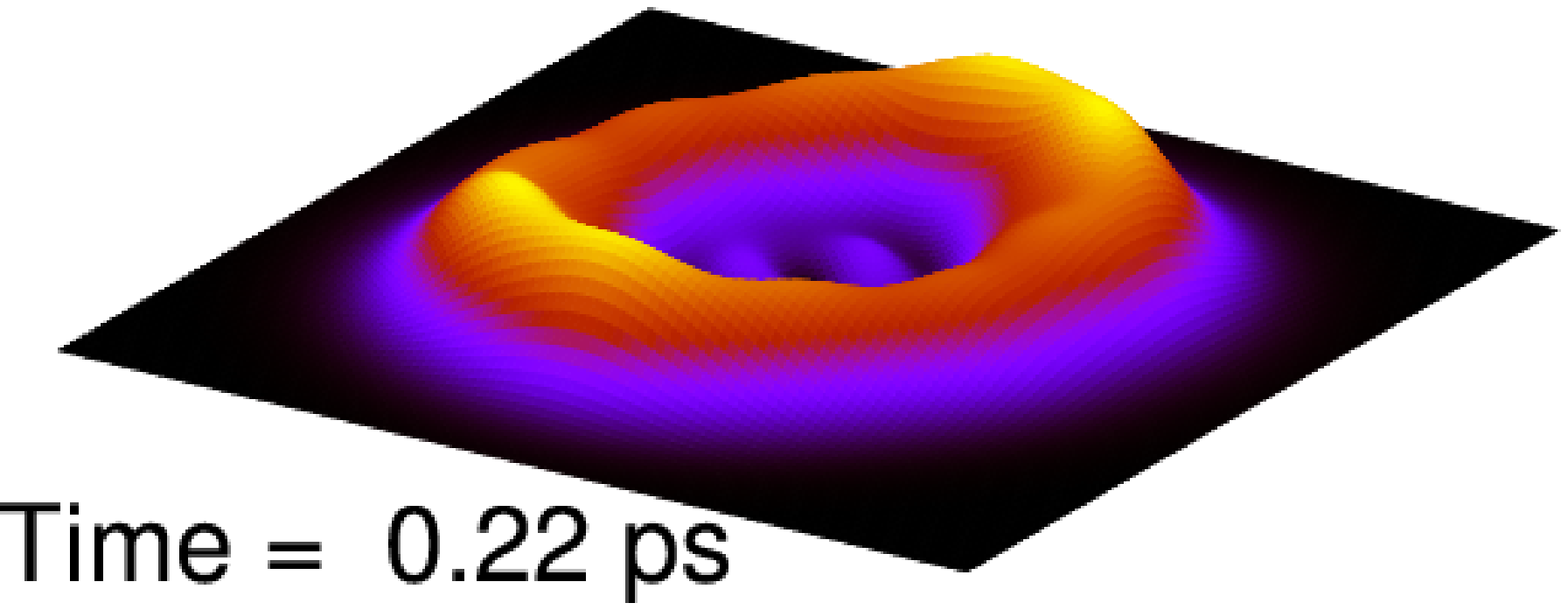}}} &
\rotatebox{270}{\scalebox{.28}{\includegraphics{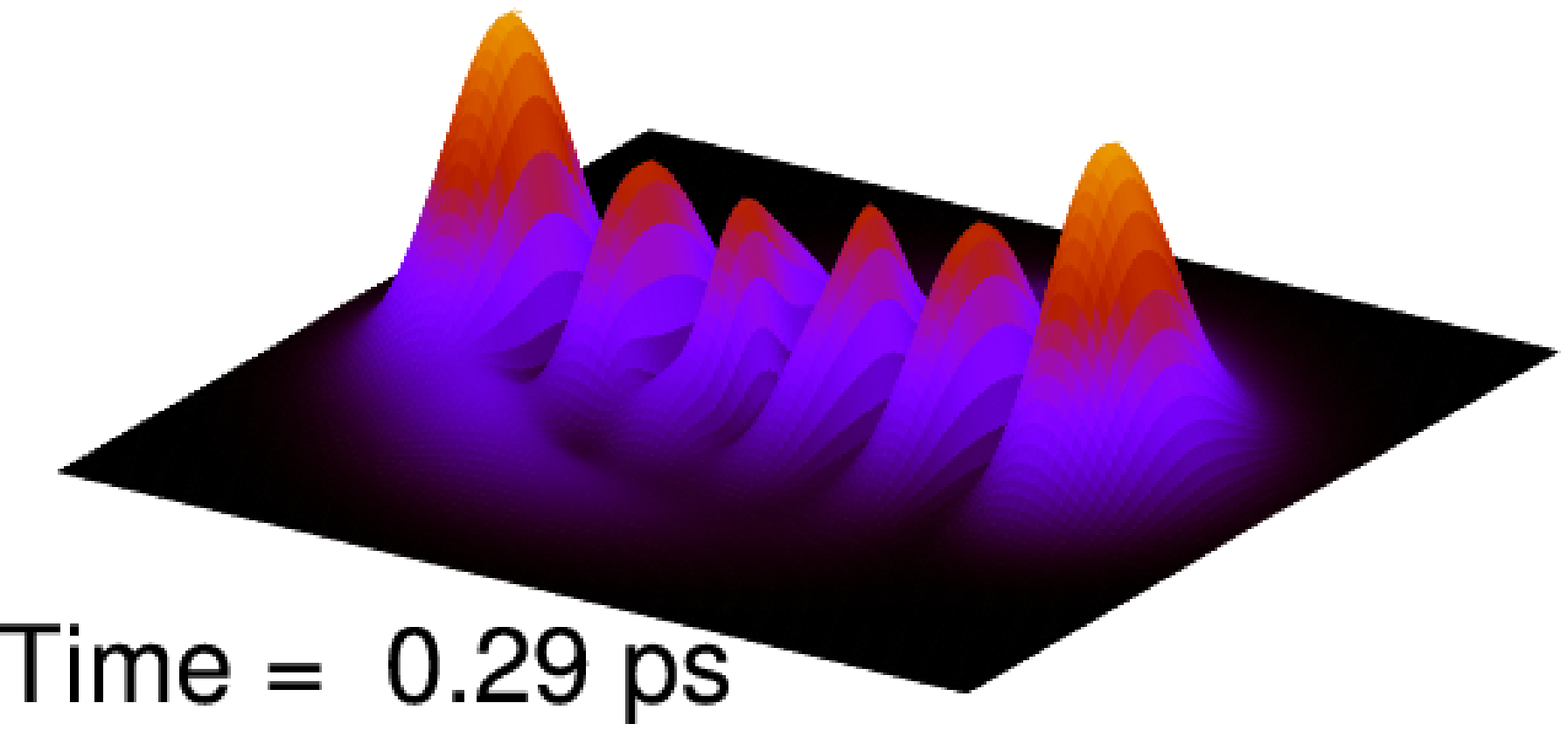}}} &
\rotatebox{270}{\scalebox{.28}{\includegraphics{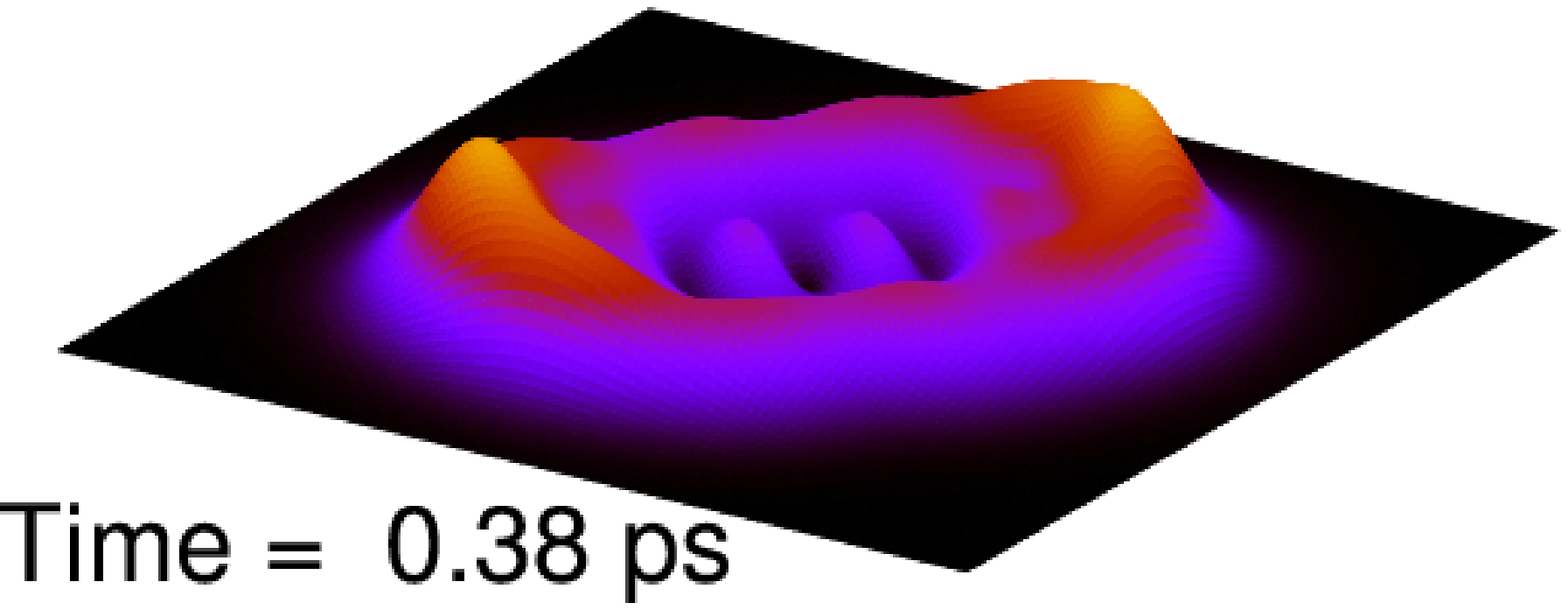}}} \\

\rotatebox{270}{\scalebox{.28}{\includegraphics{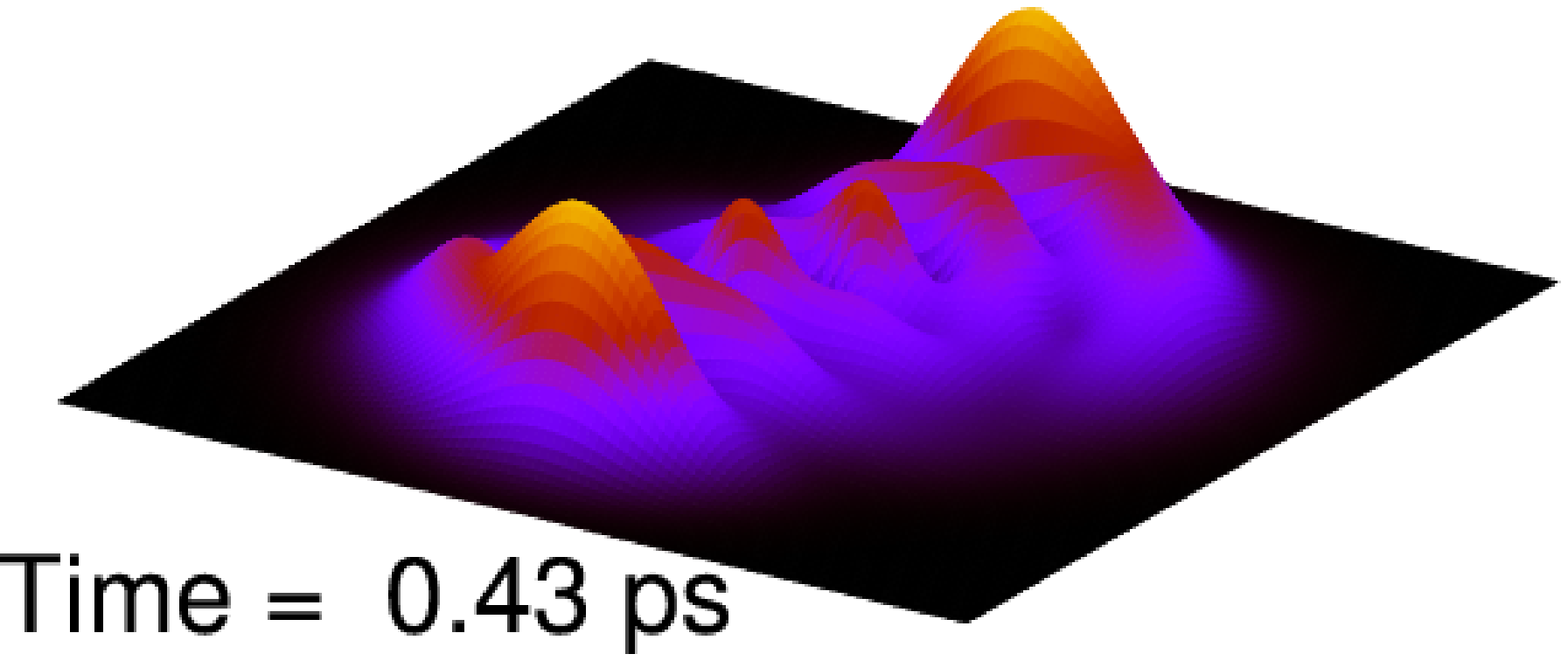}}} &
\rotatebox{270}{\scalebox{.28}{\includegraphics{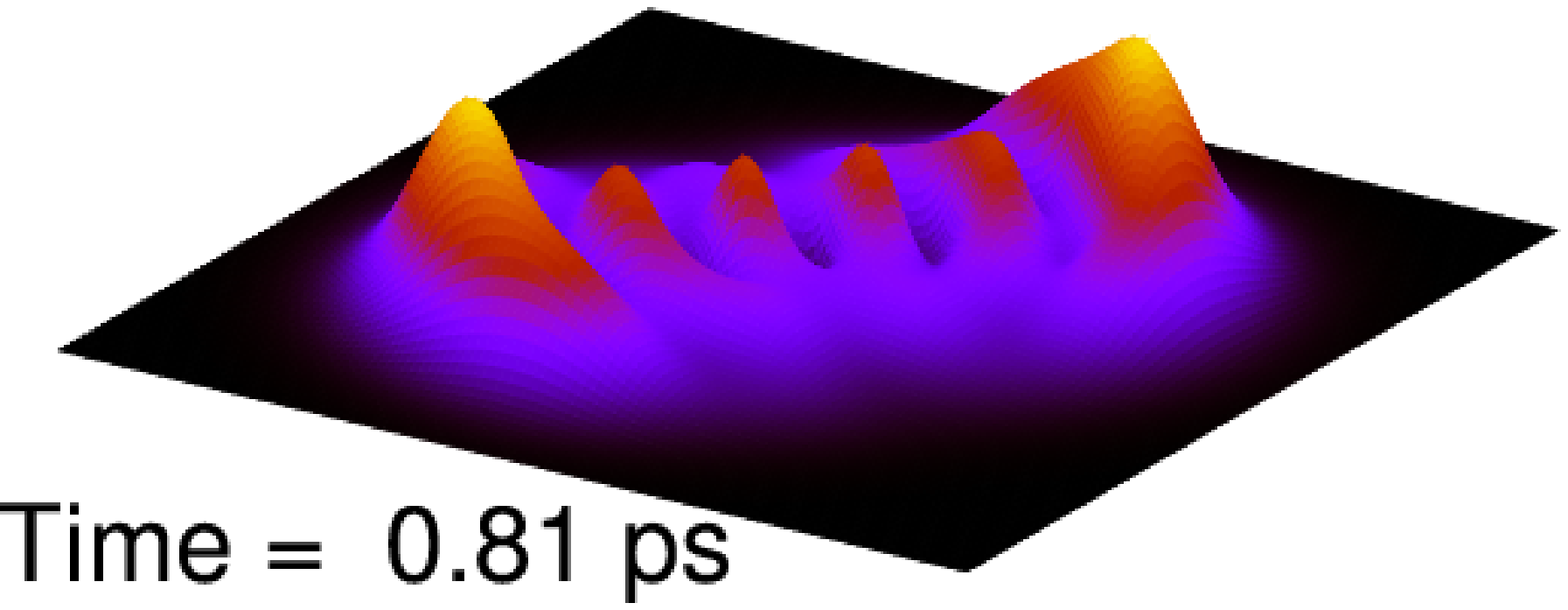}}} &
\rotatebox{270}{\scalebox{.28}{\includegraphics{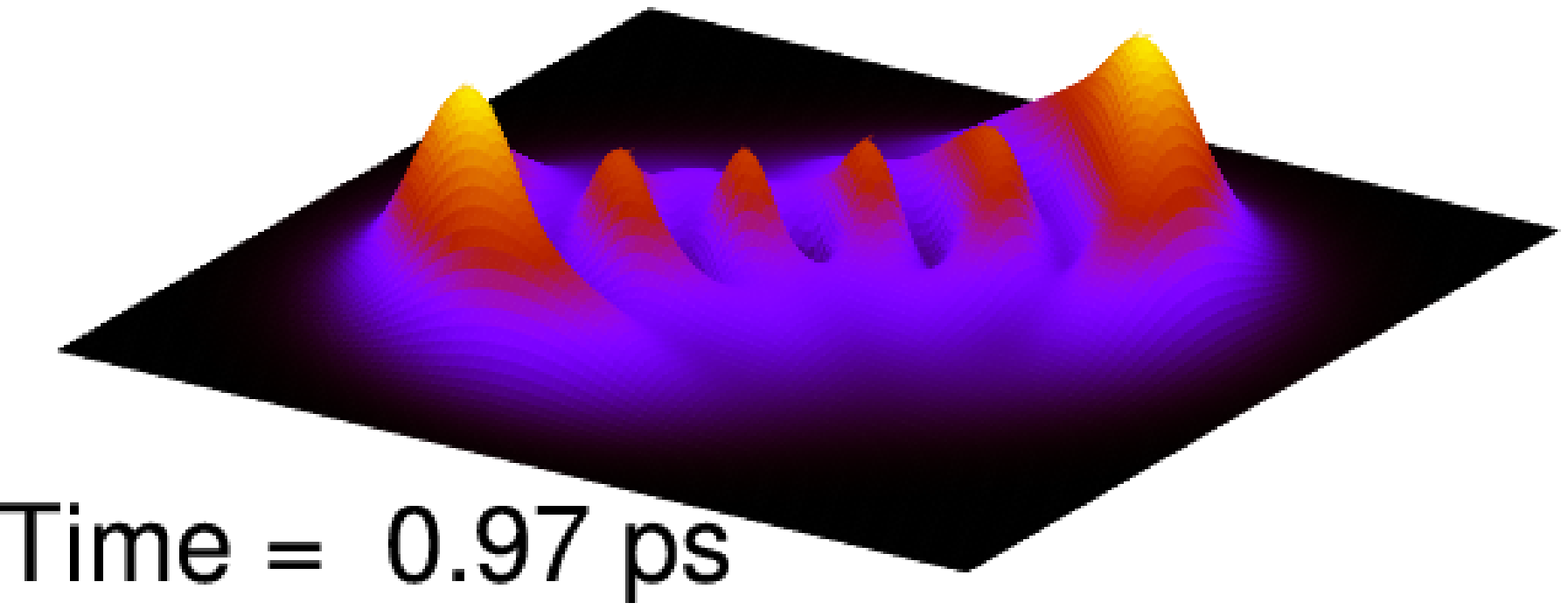}}}
\end{tabular}
\caption{Snap-shots of the time evolution of the total state density for an initial local mode state.  Oscillations between local mode states are seen at short time, but density quickly begins to obey a Boltzmann like distribution with fluctuations.\label{fig:LM}}
\end{figure*}

\begin{figure*}[htb]
\begin{tabular}{ccc}
\rotatebox{270}{\scalebox{.28}{\includegraphics{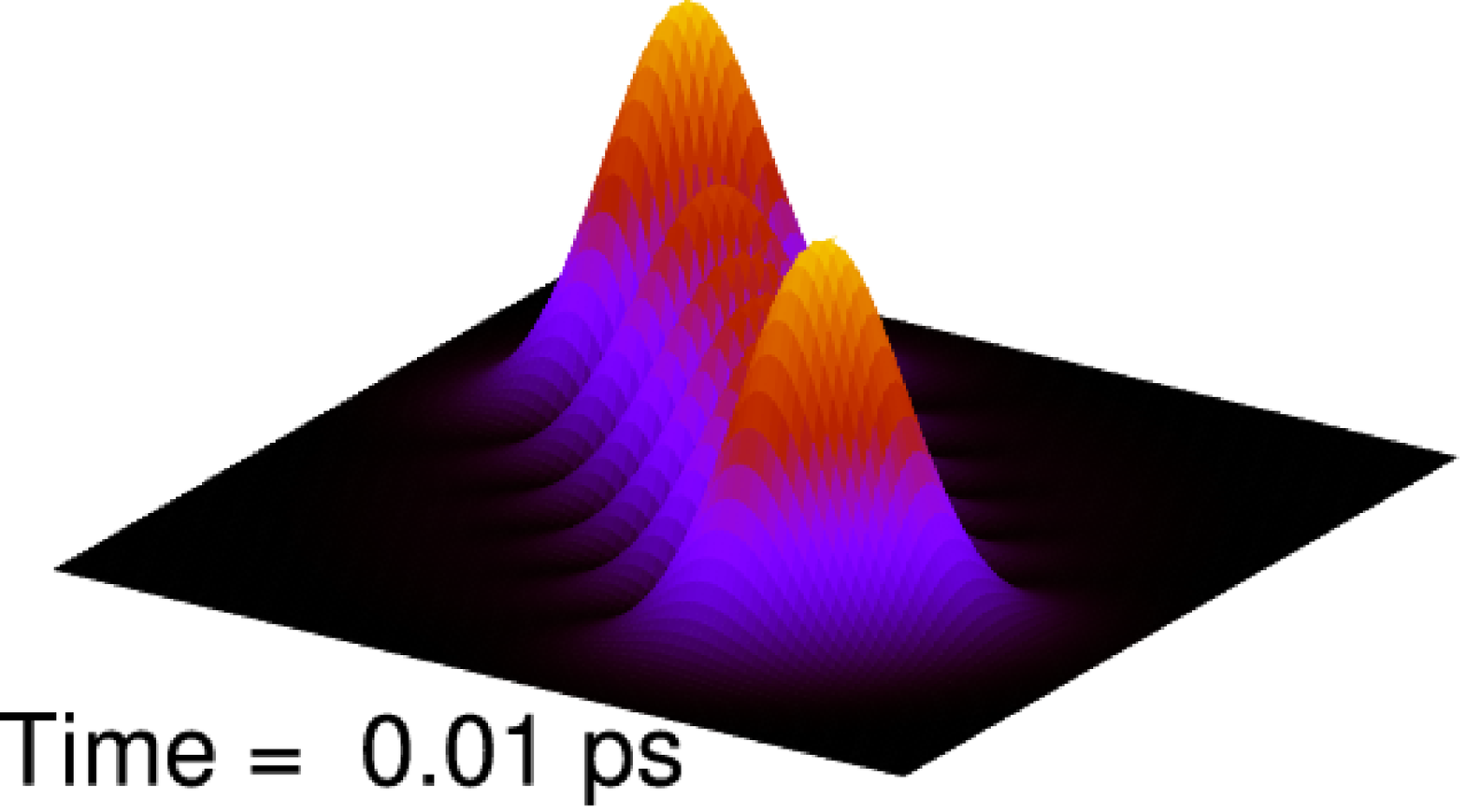}}} &
\rotatebox{270}{\scalebox{.28}{\includegraphics{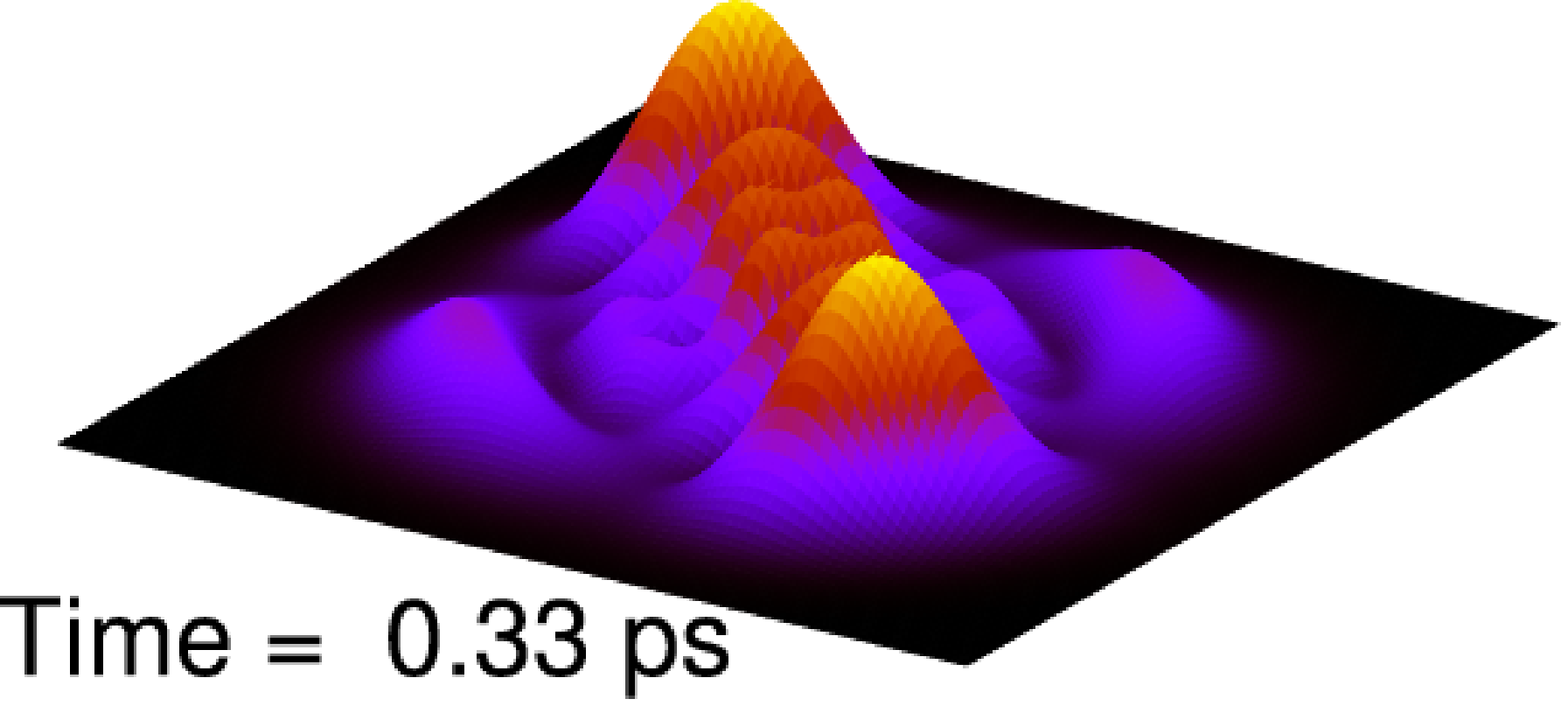}}} &
\rotatebox{270}{\scalebox{.28}{\includegraphics{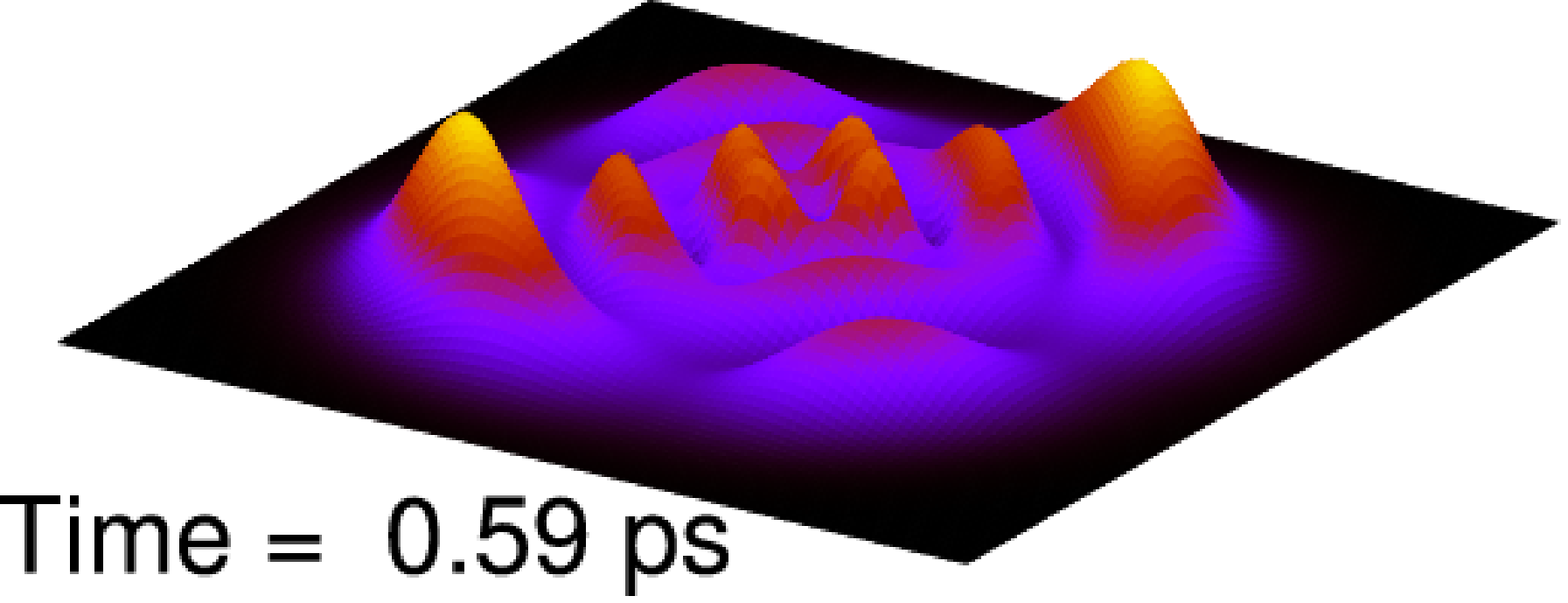}}} \\

\rotatebox{270}{\scalebox{.28}{\includegraphics{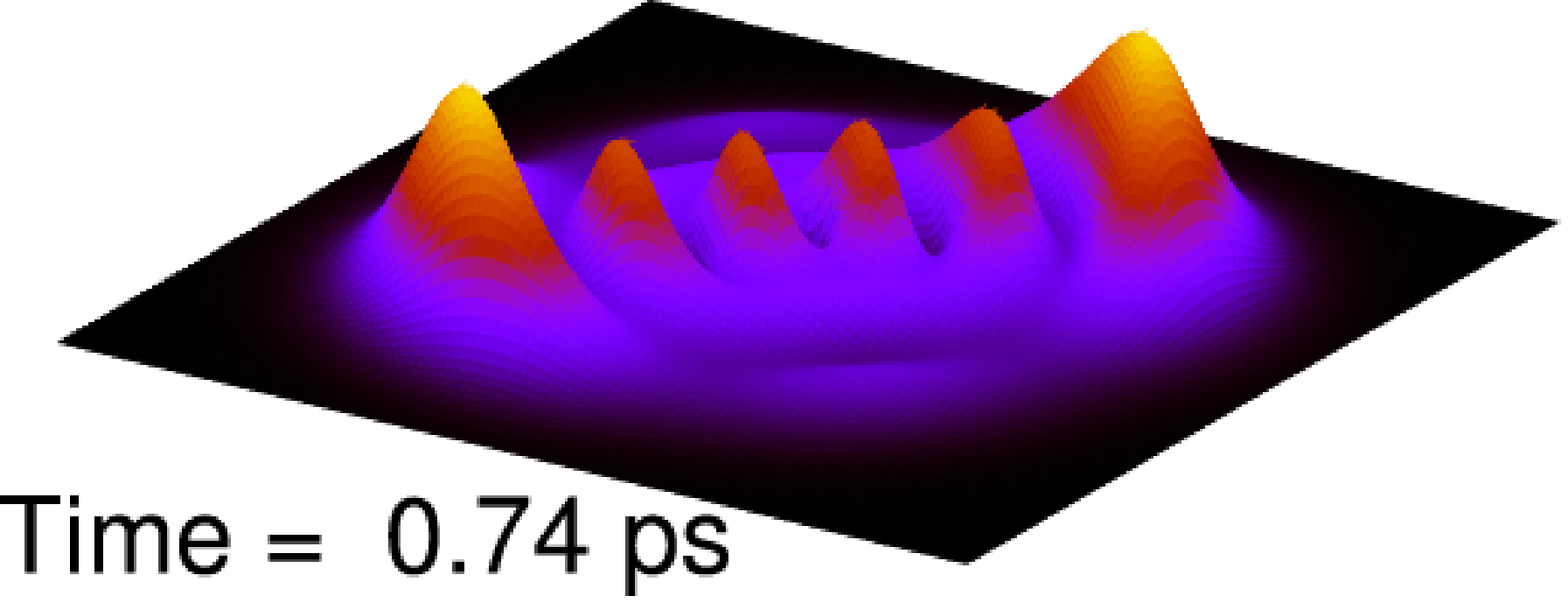}}} &
\rotatebox{270}{\scalebox{.28}{\includegraphics{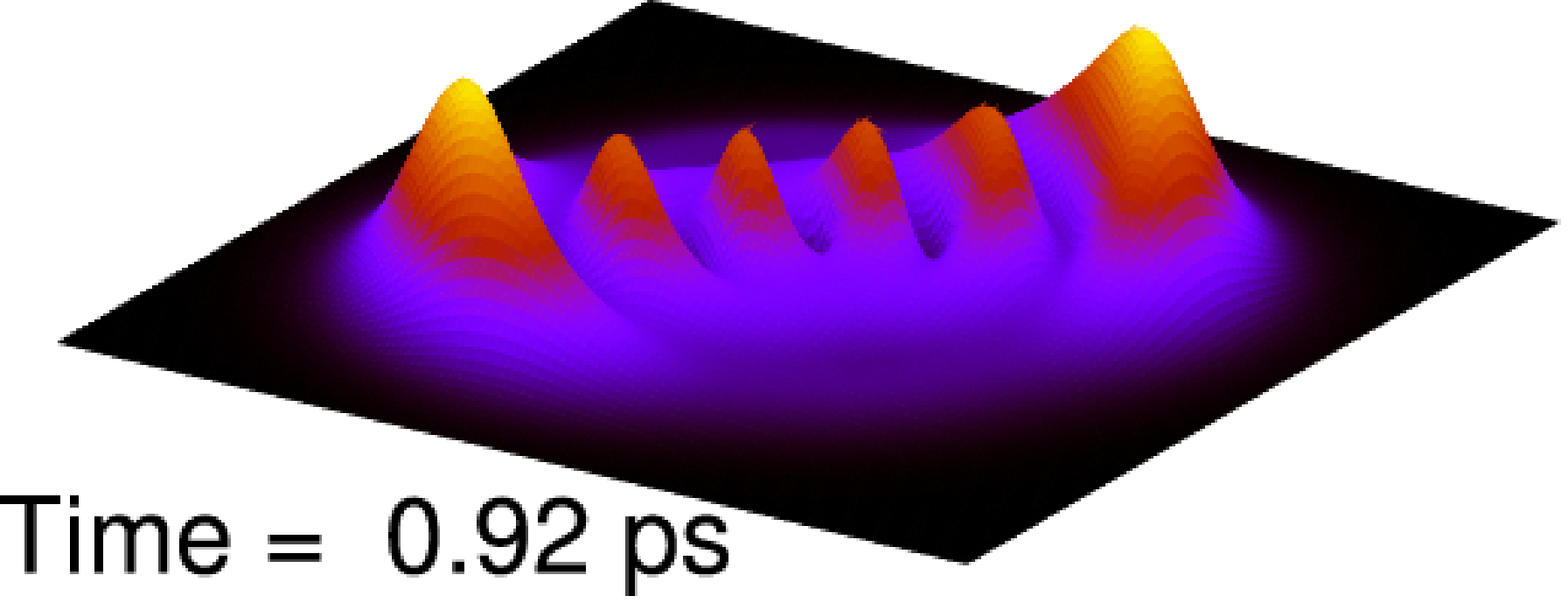}}} &
\rotatebox{270}{\scalebox{.28}{\includegraphics{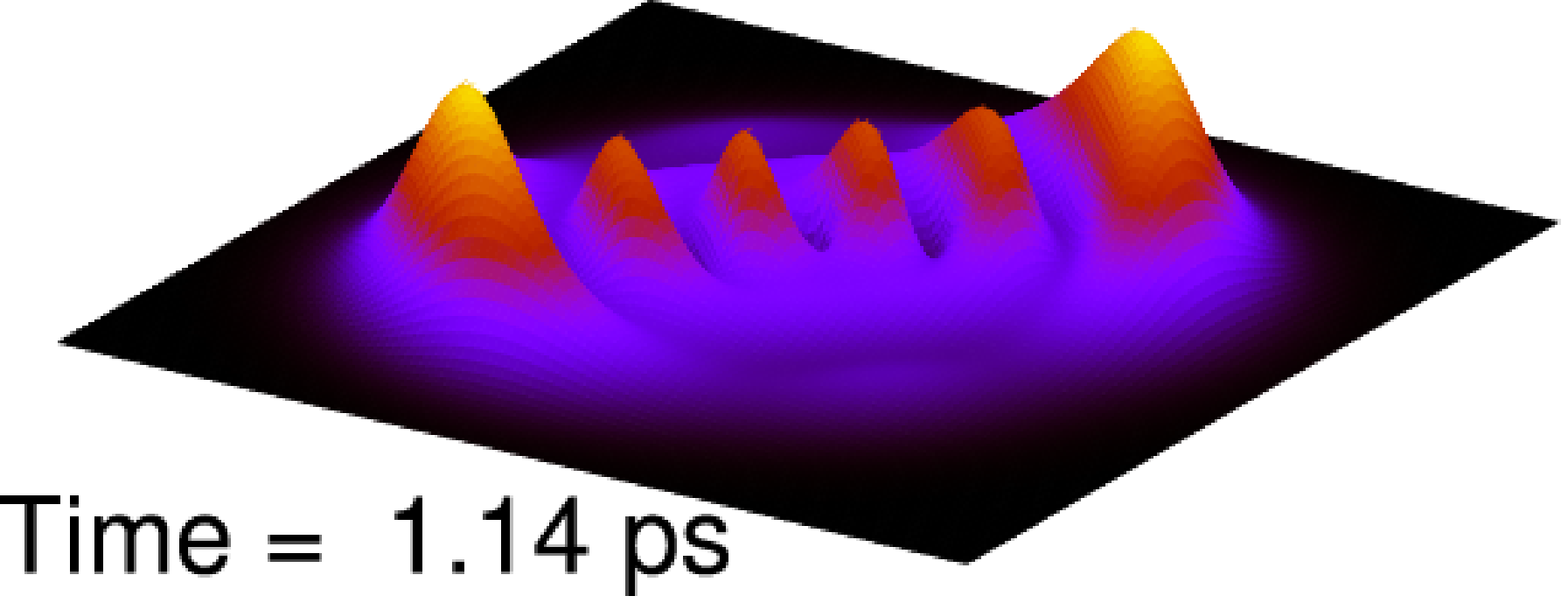}}} \\
\end{tabular}
\caption{Snap-shots of the time evolution of the total state density for the sixth level of the $N=5$ polyad.\label{fig:NSL6}}
\end{figure*}

At each instant in time the populations obtained from the RDM can be used to construct the time-dependent  spatial density of S (the coordinate representation of the density of S).  This representation of the simulation data is less abstract in some ways than the entropy and of clear use for work with realistic coordinate-dependent systems.  Snapshots of the spatial density at various points in time are shown in Figures \ref{fig:NSL1}, \ref{fig:LM}, and \ref{fig:NSL6}, which correspond to the  initial states of $n=0$, the local mode, and $n=5$, respectively.  They are ordered according to their energy expectation value.  The time evolution of the density can be complicated, as in Figure \ref{fig:LM}, or relatively simple, as in Figure \ref{fig:NSL1}.  It is interesting to note that relatively small changes in spatial density actually correspond to large changes in entropy.  For example, comparing time frames 0.01 and 0.29 ps of Figure \ref{fig:LM} shows a fairly similar spatial density, yet an examination of Figure \ref{fig:entropy} shows that over this same time period the entropy is increasing more rapidly than at any other point in the simulation.  This reveals that a simple examination of entropy or spatial density alone does not show the full story and that both ideally should also be considered when possible.

With the above observations in mind, it is instructive to consider the time evolution of the isolated S, i.e. without the SE interaction.  The energy eigenlevels of S obviously  exhibit only trivial phase evolution.  Conversely, the local mode state is a time dependent state with or without E.  Without E this state will oscillate between the two symmetry-related local modes forever.  With the SE interaction, this behavior is shown initially, although slightly modified due to E, in the first few frames of Figure \ref{fig:LM}.  As time progresses this oscillation is completely damped out by E.  At first this effect appears as a slight ``smearing'' of density. When watching movies of all time frames calculated, this smearing could also be described as ``friction" acting on the spatial density.    This eventually leads to the replacement of the oscillation between the local modes with fluctuations about the equilibrium Boltzmann density.  

 The time-independent S eigenlevels do not have any spatial oscillations to damp out, but E  does act as a driving force away from the initial spatial density.  This is strikingly illustrated in Figure  \ref{fig:NSL6} for the n = 5 antisymmetric stretch state, the highest energy member of the polyad.  The initial density collapses away from the $n=5$ state directly towards the $n=0$ state as it is the predominate level populated in the Boltzmann distribution at this temperature.  This process is not at all random.  The density does not haphazardly reach the final Boltzmann distribution but rather is directly driven towards this final spatial density, as can be appreciated by examination of the density plots.  This is striking given that nothing was done to explicitly include a directed driving force in the random SE interaction.    Initially the random SE couplings serve to drive the initial state to equilibrium.  After this is accomplished the couplings are of course still present, but now serve to create  fluctuations about the equilibrium spatial density.  On the other hand, the $n = 0$ symmetric stretch state in Figure \ref{fig:NSL1} has an initial density that points along the predominant direction of the final Boltzmann distribution, so the driving force is the damping of the quantum peaks of the initial distribution.  

It is interesting to compare the time scale of dynamics of the various states.  Examining the entropy (Figure \ref{fig:entropy}) is very useful.    If E was not present, or did not interact with S, all states would have zero entropy for all times since they would all stay as pure states.  We see that E induces an increase in entropy at a comparable rate for all initial states considered, with the local mode state falling in the middle of the envelope defined by these states.  The ``slowest'' state is the $n=0$ state while the ``fastest'' states are the $n=4$ and $n=5$ states.  Again, the most striking feature is how all states decay to the Boltzmann distribution at a very similar rate, within approximately a picosecond for each state.  The is true even for the symmetric $n = 0$ state of Figure \ref{fig:NSL1} which has an initial distribution pointed along the same direction as the Boltzmann distribution, whose largest component in the RDM is the $n = 0$ contribution.  Another interesting feature is that some of the states display an initial peak in the entropy, e.g. the $n = 5$ state.  This seems to come about because such states need to have their initial density ``squashed" while the Boltzmann-like density is growing in.  

At this point we emphasize that the final state of S is represented by an approximately diagonal RDM -- nearly full dephasing has taken place.  Given the finite size of both S and E in the calculation, the states are not perfectly dephased.   Since the Boltzmann distribution is known, it is trivial to obtain  exact populations for each S eigenlevel corresponding to a given temperature. This allows us to us obtain the perfectly dephased results for a given temperature analytically.    The state densities can then be constructed using these analytic populations, which we show in Figure \ref{fig:Bwf} for a range of temperatures.  Note that in constructing these plots we have assumed in the Hamiltonian that only the 6 states of the N=5 polyad levels are relevant and that the population cannot ``leak'' out to levels of other polyads.  At absolute zero the state density is identical to the ground state.  At the infinite temperature limit, where all populations are equal, we obtain a state density that is isotropic about the Z-axis, but shows a strong radial dependence that one might believe is indicative of quantum interference.  But as we will now show, quantum interference vanishes  at infinite temperature.  

\section{Quantum interference and von Neumann entropy}  \label{interference}

Here we present a discussion of the connection between quantum interference, coherence in the density matrix, and the von Neumann entropy, in particular what it means to have a lack of quantum interference.  Contrary to a perhaps commonly held understanding, complete dephasing  or `` loss of coherence"  in the RDM  -- such as found in a thermal distribution -- does not in general require the complete absence of quantum interference effects in the system.  

One effect of entanglement is to ``extrude" quantum superposition effects out of S into the combined SE universe.  This is one reason entanglement plays such an important role in the formation of a classical world and the quantum measurement process.  The effect of entanglement on the system is described by the reduced density matrix.    The diagonal form of the RDM gives the probabilities of the different densities that make up a mixture, as expressed in the Schmidt basis for the SE universe \cite{nielsenchuang}.   The von Neumann entropy obtained from the diagonal RDM from Eq. \ref{vNentropy}  is much like  a  classical system entropy.  The von Neumann entropy is a measure of ``mixedness" in S.   It is also a measure of quantum interference effects in S, as we shall see. Maximal entropy occurs when all the $P_i$ are equal, like the standard microcanonical ensemble.  In this situation, quantum interference effects vanish.  This  can be demonstrated as follows.

Consider a pure S state $| \psi \rangle$ expressed as a superposition in a basis $ \{ | i \rangle \} $;  and a reference (measurement) basis $ \{ | \alpha \rangle\}$.  To explore interference effects in terms of the S basis for a reference state $ | \alpha \rangle$ we examine 

\begin{equation}  | \langle \alpha | \psi \rangle |^2  = |   \sum_i c_i \langle \alpha  | i   \rangle |^2 =  \sum_{ij} c_i^* c_j \langle \alpha  | i  \rangle \langle j  | \alpha \rangle  \label{interferenceterms}    \end{equation}  

\noindent The terms $\langle \alpha | i \rangle \langle j  | \alpha  \rangle$  can be thought of as quantum ``interference" or ``coherence"  terms {\it with respect to the basis} $ \{ | i \rangle \}   $.  The terms ``interference" and ``coherence" are widely used, but not in a terribly precise manner. 

Now we consider interference terms of the type just considered as they arise for a reduced density operator $ \rho_S$  for a system interacting with an environment.   In the $d$-dimensional basis $ \{ |  \phi_i \rangle \} $ that diagonalizes this operator we have 

\begin{equation}  \rho_S = \sum_i^d P_i  | \phi_i \rangle  \langle \phi_i   |  \end{equation} 

\noindent  The density in the reference state $ | \alpha \rangle $ is calculated as 

\begin{equation}  \langle \alpha | \rho_S | \alpha \rangle = \sum_i P_i \langle \alpha | \phi_i \rangle \langle \phi_i | \alpha \rangle  \label{densitypurestate} \end{equation} 

\begin{figure}[htb]
\begin{tabular}{c}
\rotatebox{270}{\scalebox{.3}{\includegraphics{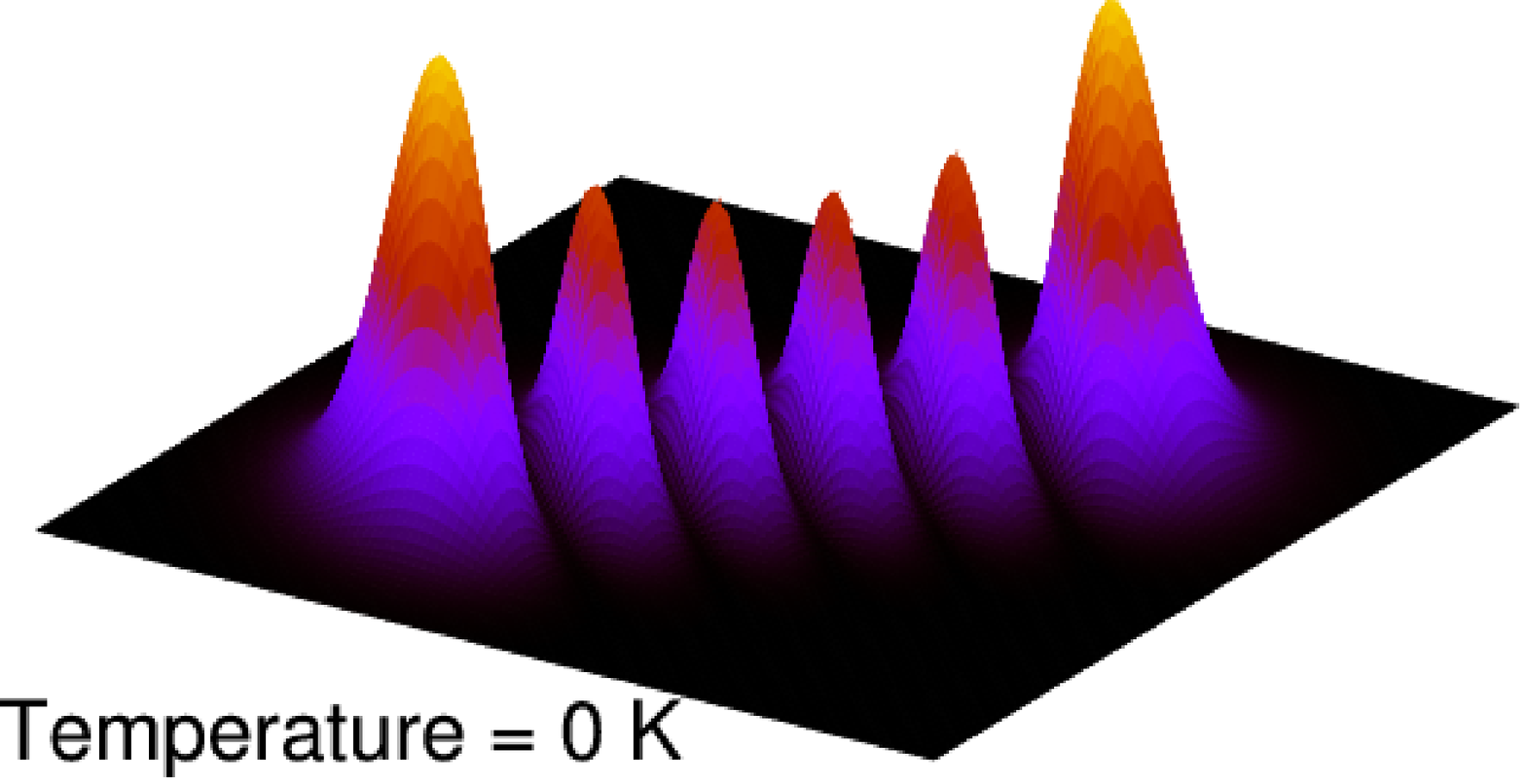}}} \\
\rotatebox{270}{\scalebox{.3}{\includegraphics{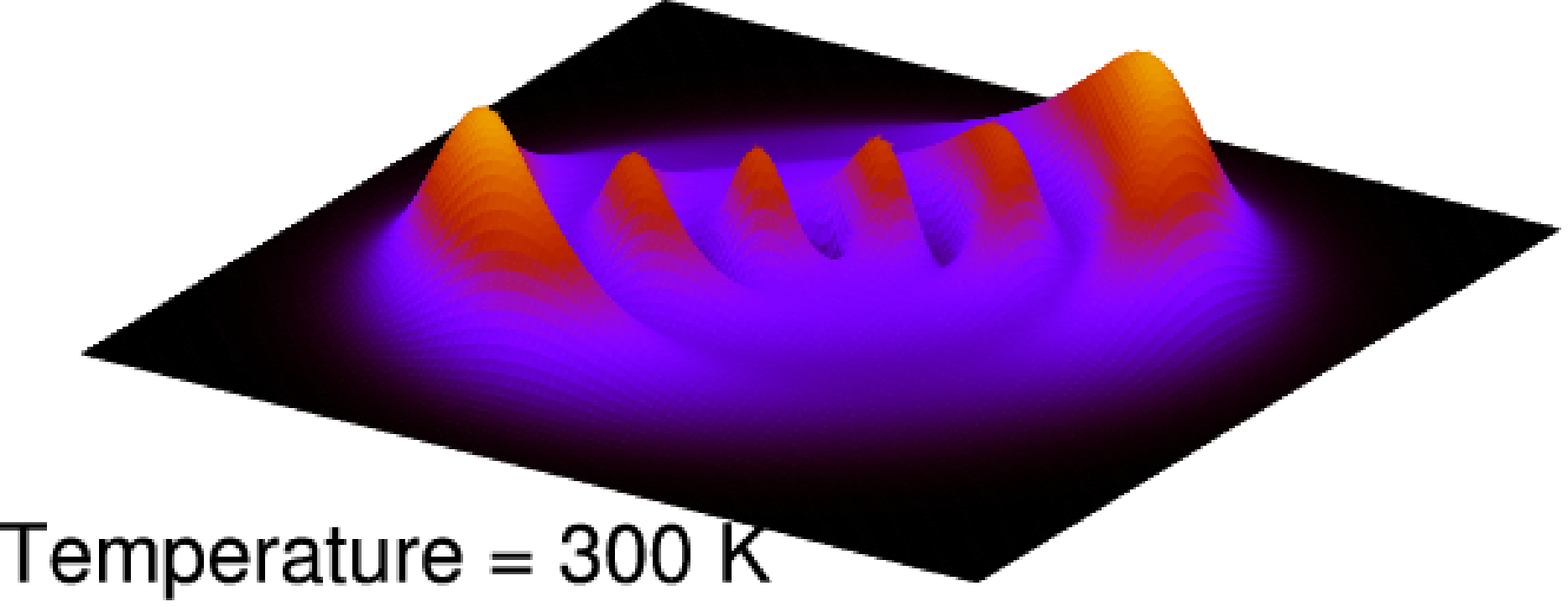}}} \\
\rotatebox{270}{\scalebox{.3}{\includegraphics{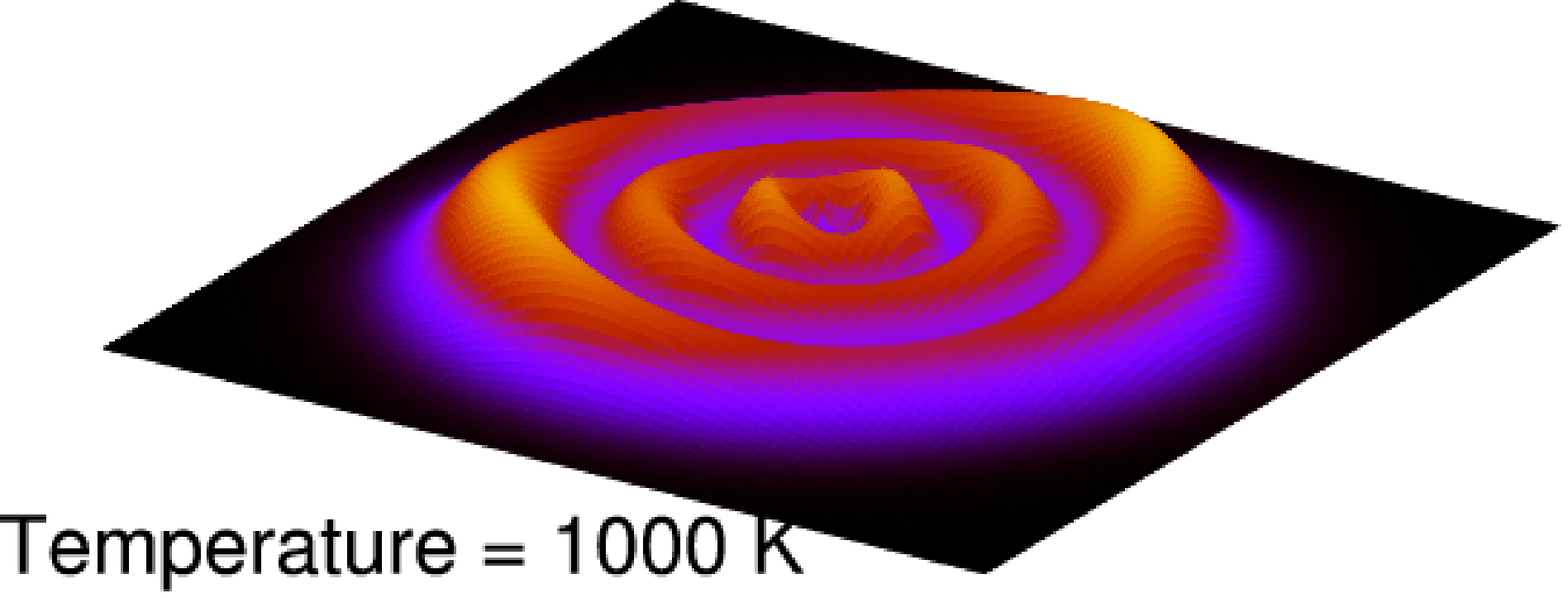}}} \\
\rotatebox{270}{\scalebox{.3}{\includegraphics{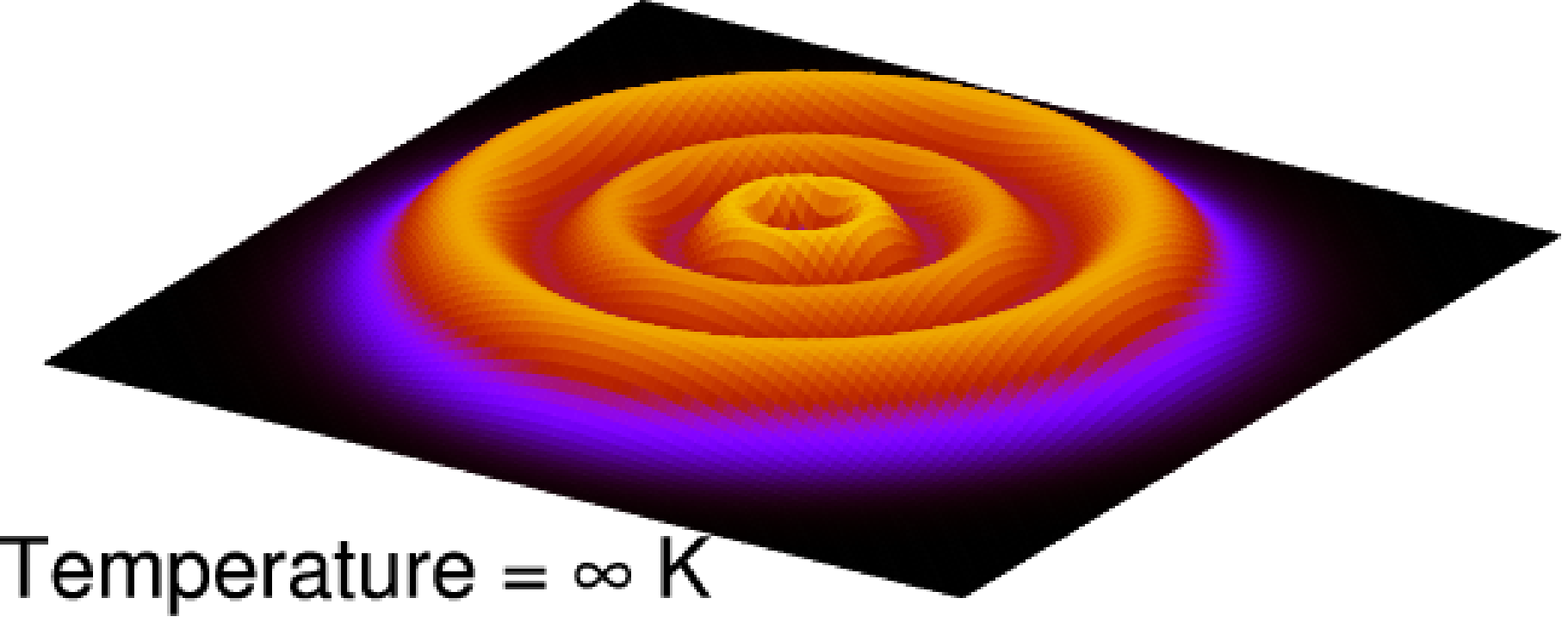}}} 
\end{tabular}
\caption{State density for various Boltzmann temperature distributions with temperature equal to 0, 300, 1000, and $\inf$ Kelvin.  We restrict the state space to the $N=5$ polyad.\label{fig:Bwf}}
\end{figure}

\noindent There are no cross terms of the kind in Eq. \ref{interferenceterms}.  However, in general there are still ``quantum interference" effects in a system with non-maximal entropy.   The system is by no means completely ``classical," even though dephasing in the off-diagonal elements of the density matrix may be complete in the given basis. These interference effects are evident in our calculations and identifiable by the anisotropic nature of the spatial density.  This holds in particular for a thermal distribution with $ P_i = e^{- E_i/kT}$ and canonical entropy:  quantum coherence is present at finite temperature, even when there is complete dephasing in the reference basis.  This is seen quite clearly in equilibrium thermalized states, e.g. for the various temperatures in Fig. \ref{fig:Bwf}.  (See discussion of the special $T = 0$ case below.)

Now suppose that $ \rho_S $ in (\ref{densitypurestate}) has maximal entropy i.e. the probabilities are all equal $P_i = 1/d$:    

\begin{equation}  \langle \alpha | \rho_S | \alpha \rangle = \sum_i  \frac{1}{d} \langle \alpha | \phi_i \rangle \langle \phi_i | \alpha \rangle  \label{diagonaldensity}    \end{equation} 

\noindent Then $\rho_S$ is proportional to the identity operator, so under a transformation to any basis $ \{ | i \rangle \} $ the matrix $\rho_S$ remains diagonal, and  the density retains the diagonal form of Eq. \ref{diagonaldensity}:

\begin{equation} \langle \alpha | \rho_S | \alpha \rangle = \sum_i  \frac{1}{d} \langle \alpha | i  \rangle \langle i | \alpha \rangle     \label{generaldiagonaldensity}              \end{equation} 

\noindent There are no cross terms of the kind in Eq. \ref{interferenceterms} {\it in any basis whatsoever}.  A way of understanding this is that under the transformation to a new basis $\{ | i \rangle \}$, cross terms appear, but in the final density calculation, all the cross terms sum to zero.  Thus, for maximal entropy i.e. all diagonal elements equal in the reduced density matrix, quantum interference terms vanish identically for all states of any reference basis.  

We do present one example of a lack of quantum interference which is not described explicitly in Equations \ref{diagonaldensity} -\ref{generaldiagonaldensity}, namely the $T=0$ Kelvin plot of Figure \ref{fig:Bwf}.  The zero temperature limit represents a special case where the state described by the RDM does not span the Hilbert space of the SE universe.  This results in a diagonal RDM basis $ \{ |  \phi_i \rangle \} $ with fewer vectors than the dimensionality of the SE universe.  In fact only one vector is required to diagonalize the RDM in the $T=0$ limit and hence there is no other state with which interference can take place.

\section{Discussion and Summary\label{sec:summary}}

The goal of this work has been to explore the time dependent dynamics of a quantum system embedded in a small quantum environment, with a view toward comparison with thermodynamic behavior.  We wished to model a system with sufficient structure to have interesting time dynamics while maintaining computational tractability.  The system of two coupled local mode oscillators, which result in a set of normal mode eigenstates, is ideal for this goal.  The time dependent dynamics of the system are revealed through the calculation of the reduced density matrix and subsequent calculation of the von Neumann entropy, system populations, and spatial density plots.

Previously Ref. \cite{Gemmerarticle}  had shown how to create an environment that resulted in a system (more primitive than ours) producing the expected eigenstate populations for a temperature $T$ for one specific system state combined with one specific environment state.  In this work, we have shown how to extend this treatment in such a way that any combination of our system and environment states will yield nearly the same final populations and hence the same temperature as would occur for a true environmental bath.  Our approach shows how to define the initial composite state of the system and environment in a way analogous to a microcanonical ensemble for a system-environment ``universe.''   This general technique can be applied to any small quantum system if the eigenvalues are available. Our results are consistent with the earlier computational findings of Gemmer et al. \cite{Gemmerarticle} and the complementary analytical work of Popescu et al. \cite{PopescuNature} that aims to show that thermalization of a quantum system that becomes entangled with a quantum environment is ``typical."

We chose to examine the time evolution of the zero order (i.e. isolated) system energy eigenstates as well as the system time dependent local mode state.  All states reach approximately the same final entropy as well as the same final fitted temperature, which is in good agreement with the temperature expected from the analytic degeneracy formula for the environment.  An examination of the time dependent dynamics of the system spatial density reveals that each state quickly loses ``memory'' of its initial density and is rapidly driven to approximately the same fluctuating equilibrium state.  The ``quantum coherences" of the off-diagonal elements of the reduced density matrix present at time zero also essentially vanish due to dephasing and population transfer.  However, contrary to common understanding, quantum coherence in the sense of ``interference effects'' vanish only with maximal von Neumann entropy, which in turn corresponds to the infinite temperature limit.  

Perhaps the most interesting result of this investigation is the phenomenon that each initial system state is driven to the same thermal equilibrium reduced density, with complete loss of memory of the initial state structure, all at essentially the same rate.  The symmetric normal mode overtone state, which looks most like the equilibrium distribution, is no different in this regard than the antisymmetric mode, which is orthogonal in coordinate space.  The same finding extends to  the local mode state, which undergoes oscillations between the two local mode overtones in the pure system initial state.  
There is no remnant of the original ``structure" of the quantum state either in the residual quantum coherence in the thermalized state, or even in the ``classical shape" that one would obtain by smoothing the density of the initial state. 

This raises interesting questions pertinent to the problem of understanding the emergence of a classical from the quantum world \cite{Schlosshauer:decoherence}.  Consider a structured object -- a rubber duck, say -- that is put in contact with a thermal bath.  The duck does not lose complete memory of its initial structure when it comes into contact with the bath.  This means that the duck system does not thermalize to a Boltzmann distribution of all the true system eigenstates.  Rather, the system is approximately constrained dynamically, so that it reveals in thermalization only a subset of states, probably of quasi-eigenstates (i.e. not true eigenstates of the unconstrained system)   that retain a very recognizable remnant of the original structure.

In the oscillator system studied here, no such ``structured" behavior is observed.  It could be that this is due to properties of the system, or of the bath, or both.   We are inclined to the view that the simple system of harmonically coupled harmonic oscillators is too simple to show  behavior of the kind just outlined.  What may be needed is a more complex, nonlinear system of the kind well-known to exhibit more complex classical phase space structure with dynamically constrained behavior.     Two-mode oscillator systems \cite{KellmanAnnRev,Xiao89sphere,Xiao90cat,Svitak95} like those considered in this paper but with nonlinear oscillators and  strong resonance couplings, such as coupled stretch normal modes in 2:2 resonance, or 2:1 Fermi resonance systems, have marked phase space structure, including stable and unstable modes born in bifurcations, and phase space barriers.  Initial system states corresponding to these different types of dynamics might plausibly be expected to have significantly different dynamics in entanglement with a quantum thermal environment, including decoherence behavior and rates of approach to equilibrium.  The results here on harmonic systems suggest that this might not be the case, but clearly detailed investigation of these nonlinear systems is warranted, especially since they correspond closely to states of real physical interest, e.g. excited molecular vibrational excitations in contact with an environment, such as in combustion or atmospheric systems.  

\noindent {\bf Acknowledgment}.  MK would like to thank David Perry for stimulating conversations about the von Neumann entropy.  This work was supported by the U.S. Department of Energy Basic Energy Sciences program under Contract  DE-FG02-05ER15634.


\end{document}